%

\ifx\mnmacrosloaded\undefined \input mn\fi

%

\newif\ifAMStwofonts

\ifCUPmtplainloaded \else
  \NewTextAlphabet{textbfit} {cmbxti10} {}
  \NewTextAlphabet{textbfss} {cmssbx10} {}
  \NewMathAlphabet{mathbfit} {cmbxti10} {} 
  \NewMathAlphabet{mathbfss} {cmssbx10} {} 
  \ifAMStwofonts
    \NewSymbolFont{upmath} {eurm10}
    \NewSymbolFont{AMSa} {msam10}
    \NewMathSymbol{\upi}     {0}{upmath}{19}
    \NewMathSymbol{\umu}     {0}{upmath}{16}
    \NewMathSymbol{\upartial}{0}{upmath}{40}
    \NewMathSymbol{\leqslant}{3}{AMSa}{36}
    \NewMathSymbol{\geqslant}{3}{AMSa}{3E}

     \let\le=\leqslant
     \let\ge=\geqslant
  \else
    \def\umu{\mu}
    \def\upi{\pi}
    \def\upartial{\partial}
  \fi
\fi


\pageoffset{0pc}{0pc}

\loadboldmathnames



\onecolumn        

\begintopmatter  

\title{On stationary oscillations of galaxies}
\author{Peter O. Vandervoort}
\affiliation{Department of Astronomy and Astrophysics, The University of
Chicago, 5640 Ellis Avenue, Chicago, IL 60637, USA}

\shortauthor{P. O. Vandervoort}
\shorttitle{Oscillations of galaxies}



\abstract {This paper describes a general investigation of
stationary oscillations of galaxies. It begins with a linear
analysis of modes of oscillation with continuous spectra of real
frequencies.  Such modes are gravitational analogues of the van Kampen
modes of oscillation in plasmas.  The characteristic value problem governing
modes of the van Kampen type in a galaxy is solved with the aid of a modified version
of the matrix method of Kalnajs in which the perturbation of the distribution function
is expressed in terms of generalized functions.  In general, there is no characteristic
equation governing the frequencies in the continuous spectrum.  However, isolated
frequencies in the continuous spectrum do satisfy a characteristic
equation which, for stellar systems, is a counterpart of the
dispersion relation proposed by Vlasov for plasma oscillations.
The linear analysis also provides a characteristic equation for
modes with a discrete spectrum of real and/or complex frequencies.
The second part of the paper describes a perturbation theory for
a stationary oscillation of a galaxy with a small but finite amplitude.
Integrals of the stellar motion are constructed with the aid of canonical
perturbation theory and used in conjunction with the theorem of Jeans
in order to specify the density of stars in the six-dimensional phase
space.  These oscillations are slightly nonlinear counterparts of the modes
of the van Kampen type, and they are stellar-dynamical counterparts
of the nonlinear plasma waves described by Bernstein, Greene, and
Kruskal.  Fully nonlinear models of stationary oscillations of galaxies
can be constructed with the aid of Schwarzschild{\rq}s numerical method
for the solution of the fundamental integral equation describing the
self-consistency of a stellar system. }


\maketitle  

\section{Introduction}

Stationary oscillations of galaxies are oscillations with constant amplitudes.
Relatively little attention has been given to stationary modes of oscillation in conventional
investigations of small perturbations of galaxies.  A major reason for this apparent
neglect is that the principal investigations of the subject have been based on methods 
which have their origin in Landau\rq{s} (1946) initial-value formulation of the theory of
plasma oscillations.  A particular feature of such methods is the adoption of the so-called
\lq{Landau contour}\rq\/ or \lq{Landau prescription}\rq\/ for the integration of singular
quantities over the momentum space (Fridman \& Polyachenko 1984; Palmer 1994).
Early investigations of the oscillations and the stability of infinite, homogeneous steller
systems (Lynden-Bell 1962; Sweet 1963) are based on Landau{\rq}s formulation of the
initial-value problem.  Landau{\rq}s prescription also underlies the matrix method formulated
by Kalnajs (1977) and used by many others ({\rm e.g.}, Polyachenko \& Shukhman 1981;
Fridman \& Polyachenko 1984; Palmer \& Papaloizou 1987; Weinberg 1989,1991a,b; Saha 1991;
Bertin et al. 1994; Palmer 1994; Robijn 1995) for investigations of instabilities in
galaxies.  Such methods are suitable for the investigation of instabilities.
However, in place of such initial-value formulations, one needs a normal-mode
analysis in order to study stationary oscillations of a galaxy within the framework
of linear perturbation theory.

The understanding and expectations that one has regarding small perturbations in
stellar systems rest heavily on what is known about plasma oscillations.  In particular,
it is well known (see, {\rm e. g.}, Jackson 1960; Clemmow \& Dougherty 1969; Ecker 1972;
Stix 1992) that the initial value formulation of Landau (1946) and the normal mode
analysis of van Kampen (1955, 1957) provide mutually complementary descriptions of
electrostatic waves in a homogeneous plasma.  The van Kampen modes with continuous spectra
of real frequencies are stationary modes of oscillation.  However, those modes are singular
in the sense that perturbations of the distribution of particles in the six-dimensional
phase space of a single particle are expressed in terms of generalized functions.  It is
frequently suggested that such distributions are artificial and would be difficult to
establish as initial conditions in a plasma.  On the other hand, the van Kampen modes
are complete (Case 1959), and they may be superposed in order to represent more physical
initial conditions.   As results of dispersion and phase mixing, such superpositions evolve
consistently with Landau{\rq}s solution of the initial value problem and exhibit
forms of apparently irreversible behavior such as Landau damping.  Conversely,
when a van Kampen mode is adopted as an initial condition, the solution of the initial
value problem in accordance with Landau{\rq}s prescription is that van Kampen mode.
Thus, the study of the van Kampen modes, the linear modes of stationary oscillation
in a plasma, contributes significantly to our understanding of plasma oscillations,
notwithstanding that such modes might be difficult to excite individually and superpostions
of such modes would suffer Landau damping.

The literature of plasma physics shows that van Kampen modes are significant and important
in other respects as well.  The van Kampen modes are the small-amplitude limits of nonlinear
plasma waves ({\lq}BGK waves{\rq}) of the kind described by Bernstein, Greene, and Kruskal
(1957; Bohm and Gross 1949; Jackson 1960; Stix 1992).  The nonlinearity of BGK waves precludes
a principle of superposition. Thus, BGK waves are stationary oscillations of a homogeneous plasma,
and they do not suffer Landau damping.  As the linear limit of a BGK wave, each van Kampen
mode must be considered individually.  The interpretation in terms a BGK wave shows how a van Kampen
mode owes its self-consistency to the way in which orbits trapped in the electrostatic potential
of the wave are populated with particles.  The connection also establishes the van Kampen modes as
a starting point in linear perturbation theory for the study of a certain class of nonlinear
waves.

For stellar systems, this paper describes an investigation of stationary oscillations
which are the gravitational counterparts of van Kampen modes and BGK waves.
Section 2 describes the construction of modes of oscillation and instability of a stellar
system within the framework of linear perturbation theory and with the aid of a modified
version of the matrix method of Kalnajs (1977; Fridman \& Polyachenko 1984; Palmer 1994).
In particular, this formulation describes singular modes with continuous spectra of real
frequencies.  The treatment unites the methods of Kalnajs and van Kampen and generalizes
both.  Section 3 describes the construction of slightly nonlinear oscillations of stellar
systems.  In order to solve the governing equations, we make use of canonical
perturbation theory (Goldstein 1980) in order to construct suitable integrals of the
motion for the stellar orbits, and we apply the theorem of Jeans in order to specify the
distribution of stars in the six-dimensional phase space.  The oscillations derived in this
way are slightly nonlinear counterparts of the singular modes of the van Kampen type that are
described in Section 2.  As in the case of BGK waves, such stationary oscillations of a galaxy
owe their self-consistency to the way in which resonant orbits are populated with stars.
The nonlinear treatment imposes certain constraints on the frequencies of modes of the van Kampen
type which seem not to arise within the framework of linear perturbation theory.  Section 4
contains a very brief discussion of fully nonlinear oscillations of stellar systems, and the
paper concludes in Section 5 with a summary and discussion of the important results of this
work.

Remarkably, many of the elements of what is fully worked out in the present investigation have been
anticipated by Louis and Gerhard (1988), who asked the question {\lq}Can galaxies oscillate?{\rq\/}
and then constructed a self-consistent model of a spherically symmetric stellar system in a
state of stationary oscillation.  They constructed that model with the aid of a suitably generalized
version of Schwarzschild{\rq}s (1979) method for the solution of the fundamental integral equation
describing the self-consistency of a stellar system.  That paper provides an important example of
a fully nonlinear (albeit numerical) model of a galaxy in a state of stationary oscillation.  The work
of Louis and Gerhard and its relationship to the present investigation are the main subjects of Section 4.
It is relevant to the motivation for the present investigation to note that Louis and Gerhard (1988)
begin their paper with a brief review of observational evidence that states of equilibrium may
not be the only {\lq}normal{\rq\/} states of galaxies.  They suggest that some observed galaxies
could be in states of stationary oscillation.

The generality of the present investigation should be emphasized.  In the first place, what follows
establishes for finite, inhomogeneous stellar systems many results which have been well established
for infinite, homogeneous plasmas and stellar systems.  Moreover, these results apply to a wide
class of stellar systems.  The only significant restriction is that we consider the stationary
oscillations of a system in which the motions of stars in the unperturbed gravitational potential
are separable.  For the sake of clarity and specificity, we include appendices in which the general
analysis described in Sections 2 and 3 is worked out explicitly for plane waves in an infinite,
homogeneous system of stars.

\section{Modes of oscillation of the van Kampen type}

Infinitesimal perturbations in a galaxy are governed by the collisionless
Boltzmann equation in the linearized form
$$
{ {\upartial f_1 } \over { \upartial t}}
 + {\bmath{v}} \cdot { {\upartial f_1 } \over { \upartial {\bmath{x}}}}
 - { {\upartial V_0 } \over { \upartial {\bmath{x}}}} \cdot 
{ {\upartial f_1 } \over { \upartial {\bmath{v}}}}
 = { {\upartial V_1 } \over { \upartial {\bmath{x}}}} \cdot 
{ {\upartial f_0 } \over { \upartial {\bmath{v}}}} , \eqno\stepeq
$$
where the distribution function $f_0 ({\bmath{x}},{\bmath{v}})$ represents
the unperturbed density of stars in the six-dimensional phase space
of a single star, $V_0 ({\bmath{x}})$ denotes the unperturbed gravitational
potential of the system, and $f_1 ({\bmath{x}},{\bmath{v}},t)$
and $V_1 ({\bmath{x}},t)$ denote the perturbations of the distribution
function and the potential, respectively.  Here, $ {\bmath{x}} $ and
$ {\bmath{v}} $ denote the position and velocity of a star, respectively,
and $t$ denotes the time.  The density of the perturbation
$$
\rho_1 ({\bmath{x}},t) = m_* \int f_1 ({\bmath{x}},{\bmath{v}},t) {\rm d} {\bmath{v}} , \eqno\stepeq
$$
where $m_*$ is the mass of a star, is the source of the potential $V_1 ({\bmath{x}},t)$.
In other words,
$$
V_1 ({\bmath{x}},t) = - G \int_V { {\rho ({\bmath{x}}^\prime , t) }
\over {|{\bmath{x}}^\prime - {\bmath{x}}|} }
 {\rm d} {\bmath{x}}^\prime , \eqno\stepeq
$$
where the integration extends over the volume $V$ of the system.

The normal modes of the system are described by solutions of
equations (1)-(3) for $f_1 ({\bmath{x}},{\bmath{v}},t)$, $\rho_1 ({\bmath{x}},t)$,
and $V_1 ({\bmath{x}},t)$ with a time dependence $\exp (- {\rm i} \sigma t)$,
where the characteristic frequency $\sigma$ is a constant.

\subsection{Solution for $f_1 (\bmath{x},\bmath{v}, t)$ in terms of action-angle variables}

We assume that the motion of a star in the unperturbed potential $V_0 ({\bmath{x}})$
is separable.  Accordingly, with the aid of a suitable canonical
transformation $({\bmath{x}},{\bmath{v}}) \rightarrow ({\bmath{w}},{\bmath{J}})$,
we introduce a set of angle variables ${\bmath{w}} = (w_1,w_2,w_3)$
and conjugate action variables ${\bmath{J}} = (J_1,J_2,J_3)$ which describe the
motion of a star {\em in the unperturbed system\/}.  When expressed as
functions of the action-angle variables, functions of  ${\bmath{x}}$ and ${\bmath{v}}$
are quasi-periodic functions of the angle variables ${\bmath{w}}$ with periods $2 \upi$.

The actions ${\bmath{J}}$ are isolating integrals of the motion of a star in the
gravitational field of the unperturbed system.  Therefore, the unperturbed
distribution function can be expressed as a function
$f_0 ({\bmath{x}},{\bmath{v}}) = f_0({\bmath{J}})$ of the actions
in accordance with the theorem of Jeans.

Transformed to action-angle variables, the linearized, collisionless Boltzmann
equation now reads
$$
{ {\upartial f_1 } \over { \upartial t}}
 + {\bmath{\omega}}({\bmath{J}}) \cdot { {\upartial f_1 } \over { \upartial {\bmath{w}}}}
 = { {\upartial V_1 } \over { \upartial {\bmath{w}}}} \cdot 
{ {\upartial f_0 } \over { \upartial {\bmath{J}}}} , \eqno\stepeq
$$
where the quantities ${\bmath{\omega}}({\bmath{J}}) = (\omega_1,\omega_2,\omega_3)$
are the frequencies of the unperturbed motion of a star.  In other words,
${\bmath{\omega}} ({\bmath{J}}) = \upartial H_0 / \upartial {\bmath{J}}$, where $H_0 ({\bmath{J}})$
is the Hamiltonian which governs the unperturbed motion.

In order to solve equation (4), we express the perturbations of the distribution function
and the gravitational potential as trigonometric series in the angle variables of the forms
$$
f_1 ({\bmath{x}},{\bmath{v}},t) = f_1 ({\bmath{w}},{\bmath{J}},t) 
= \sum_{\bmath{n}} f_1 ({\bmath{n}},{\bmath{J}},\sigma)
\exp ({\rm i} {\bmath{n}} \cdot {\bmath{w}} - {\rm i} \sigma t) \eqno\stepeq
$$
and
$$
V_1 ({\bmath{x}},t) = V_1 ({\bmath{w}},{\bmath{J}},t) 
= \sum_{\bmath{n}} V_1 ({\bmath{n}},{\bmath{J}},\sigma)
\exp ({\rm i} {\bmath{n}} \cdot {\bmath{w}} - {\rm i} \sigma t) , \eqno\stepeq
$$
respectively, where the sums extend over sets of integers ${\bmath{n}} = (n_1,n_2,n_3)$.
In this representation, trigonometric components of equation (4) separate, and the coefficients
$f_1 ({\bmath{n}},{\bmath{J}},\sigma)$ and $V_1 ({\bmath{n}},{\bmath{J}},\sigma)$ satisfy the equation
$$
-{\rm i} [\sigma - {\bmath{n}} \cdot \bmath{\omega} (\bmath{J})] f_1 ({\bmath{n}},{\bmath{J}},\sigma)
= {\rm i} {\bmath{n}} \cdot { {\upartial f_0 } \over { \upartial {\bmath{J}}}}
V_1 ({\bmath{n}},{\bmath{J}},\sigma) . \eqno\stepeq
$$

The solution of equation (7) for $f_1 ({\bmath{n}},{\bmath{J}},\sigma)$ must allow the integration of
quantities involving $f_1(\bmath{x},\bmath{v},t)$ over the entire region of the phase space that
is accessible to stars in the unperturbed system.  In particular, $f_1 ({\bmath{n}},{\bmath{J}},\sigma)$
must be defined for this purpose at values of $\bmath{J}$ such that
$\sigma - {\bmath{n}} \cdot \bmath{\omega} (\bmath{J}) = 0$.
For each set of integers $\bmath{n}$, that condition defines a surface of two dimensions in the
three-dimensional space of the actions $\bmath{J}$.  Let two variables
$(\varpi_1,\varpi_2) = {\bmath{\varpi}}$, say, label points on that surface.  Then the values of
the actions at a given point $\bmath{\varpi}$ on the surface are given by the functions
$\bmath{J} = \bmath{J}_R (\bmath{n},\bmath{\varpi},\sigma)$ (say).

The solution for $f_1 (\bmath{n},\bmath{J},\sigma)$ is a generalized function of the form
$$
f_1 (\bmath{n},\bmath{J},\sigma) = -P { {V_1 ({\bmath{n}},{\bmath{J}},\sigma)} \over
{\sigma - {\bmath{n}} \cdot \bmath{\omega} (\bmath{J})} }
{\bmath{n}} \cdot { {\upartial f_0 } \over { \upartial {\bmath{J}}}}
+ \int \lambda (\bmath{n},\bmath{\varpi},\sigma) 
\delta[ \bmath{J} - \bmath{J}_R (\bmath{n},\bmath{\varpi},\sigma) ] {\rm d} \bmath{\varpi} ,
\eqno\stepeq
$$
where $P$ signifies that the integral of a quantity must be interpreted as
the Cauchy principal value, $\lambda (\bmath{n},\bmath{\varpi},\sigma)$
is an arbitrary function, and $\delta$ denotes Dirac{\rq}s delta function.
The integral on the right-hand side of equation (8) extends over the entire surface in
the action space on which $\sigma - {\bmath{n}} \cdot \bmath{\omega} (\bmath{J}) = 0$.
Equation (8) expresses $f_1 (\bmath{n},\bmath{J},\sigma)$ as a sum of a particular solution
of equation (7) and a general solution of the reduced equation
$[\sigma - {\bmath{n}} \cdot \bmath{\omega} (\bmath{J})] f_1 ({\bmath{n}},{\bmath{J}},\sigma) = 0$.
In particular, $\lambda (\bmath{n},\bmath{\varpi},\sigma) 
\delta[ \bmath{J} - \bmath{J}_R (\bmath{n},\bmath{\varpi},\sigma) ] {\rm d} \bmath{\varpi}$ is
the contribution to the solution of the reduced equation (for fixed values of $\bmath{n}$ and
$\sigma$) at values of the actions in the neighborhood of the values
$\bmath{J} = \bmath{J}_R (\bmath{n},\bmath{\varpi},\sigma)$.  The integration over $\bmath{\varpi}$
adds such contributions over all points $\bmath{J}$ on the surface
$\sigma - {\bmath{n}} \cdot \bmath{\omega} (\bmath{J}) = 0$.

Equation (8) clearly satisfies equation (7) where
$\sigma - {\bmath{n}} \cdot \bmath{\omega} (\bmath{J}) \ne 0$ and
$\bmath{J} \ne \bmath{J}_R (\bmath{n},\bmath{\varpi},\sigma)$.  Where
$\sigma - {\bmath{n}} \cdot \bmath{\omega} (\bmath{J}) = 0$, equation (8)
also satisfies equation (7) in the sense that
$$
\eqalign{ & \{\sigma - {\bmath{n}} \cdot \bmath{\omega} [\bmath{J}_R (\bmath{n},\bmath{\varpi},\sigma)]\}
f_1 [{\bmath{n}},\bmath{J}_R (\bmath{n},\bmath{\varpi},\sigma),\sigma] \cr
&= \lim_{\bmath{J} \to \bmath{J}_R (\bmath{n},\bmath{\varpi},\sigma)}
\Big\{ [\sigma - {\bmath{n}} \cdot \bmath{\omega} (\bmath{J})]
\Big[ -  { {V_1 ({\bmath{n}},{\bmath{J}},\sigma)} \over
{\sigma - {\bmath{n}} \cdot \bmath{\omega} (\bmath{J})} }
{\bmath{n}} \cdot { {\upartial f_0 (\bmath{J}) } \over { \upartial {\bmath{J}}}} \Big]
+  \int \lambda (\bmath{n},\bmath{\varpi}^\prime,\sigma) [\sigma - {\bmath{n}} \cdot \bmath{\omega} (\bmath{J})]
\delta[ \bmath{J} - \bmath{J}_R (\bmath{n},\bmath{\varpi}^\prime,\sigma) ]
{\rm d} \bmath{\varpi}^\prime \Big\} \cr
& = - \Big[ {\bmath{n}} \cdot { {\upartial f_0 (\bmath{J}) } \over { \upartial {\bmath{J}}}}
V_1 ({\bmath{n}},{\bmath{J}},\sigma) \Big]_{\bmath{J} = \bmath{J}_R (\bmath{n},\bmath{\varpi},\sigma)}
 . \cr } \eqno\stepeq
$$
Here the integral over $\bmath{\varpi}$ vanishes in the limit, because
$$
\sigma - {\bmath{n}} \cdot \bmath{\omega} (\bmath{J}) \to
- [ \bmath{J} - \bmath{J}_R (\bmath{n},\bmath{\varpi}^\prime,\sigma) ] \cdot
\Big\{ {\upartial \over {\upartial \bmath{J}}}[{\bmath{n}} \cdot \bmath{\omega} (\bmath{J})]
\Big\}_{\bmath{J} = \bmath{J}_R (\bmath{n},\bmath{\varpi}^\prime,\sigma)}
\quad \hbox{ as } \quad
\bmath{J} \to \bmath{J}_R (\bmath{n},\bmath{\varpi}^\prime,\sigma) . \eqno\stepeq
$$

Substituting from equations (8) for the coefficients $f_1 (\bmath{n},\bmath{J},\sigma)$ in
equation (5) we obtain the solution for the perturbation of the distribution function in
the form
$$
\eqalign{ f_1 ({\bmath{x}},{\bmath{v}},t) =& f_1 ({\bmath{w}}, {\bmath{J}},t) \cr
= - & \sum_{\bmath{n}} P { {V_1 ({\bmath{n}},{\bmath{J}},\sigma)} \over 
{\sigma - {\bmath{n}} \cdot {\bmath{\omega}} ({\bmath{J}})}} 
{\bmath{n}} \cdot {{\upartial f_0} \over {\upartial {\bmath{J}}}}
 \exp ({\rm i} {\bmath{n}} \cdot {\bmath{w}} - {\rm i} \sigma t) 
+ \sum_{\bmath{n}} \int {\rm d} {\bmath{\varpi}} \lambda ({\bmath{n}}, {\bmath{\varpi}}, \sigma)
\delta [{\bmath{J}} - {\bmath{J}}_R ({\bmath{n}} ,{\bmath{\varpi}},\sigma)]
\exp ({\rm i} {\bmath{n}} \cdot {\bmath{w}} - {\rm i} \sigma t) . \cr} 
\eqno\stepeq
$$

It is in the solution of equation (7) that the present formulation of the matrix method
differs from the formulation by Kalnajs (1977).  Where the solution by Kalnajs is based on
Landau{\rq}s (1946) initial-value analysis of plasma oscillations, the present solution is based on
van Kampen{\rq}s (1955) normal-mode analysis.  The difference is manifest in the interpretation of
equations (8) and (11) in terms of Cauchy principal values of integrals and in the appearance in those
equations of Dirac{\rq}s delta function.  Thus, for example, equation (11) is to be contrasted with
equation (10) in Section 4 of the Appendix of Fridman \& Polyachenko (1984) or with equations
(9.5) and (9.6) in Palmer (1994).

\subsection{A matrix method for the solution of the characteristic value problem
governing normal modes}

Substituting from equation (11) for $f_1 ({\bmath{x}},{\bmath{v}},t)$ in equation (2) and
rearranging terms in the resulting equation, we obtain
$$
\eqalign{ \rho_1 ({\bmath{x}},t) + 
\sum_{\bmath{n}} m_* P \int {\rm d} \bmath{v} { {V_1 ({\bmath{n}},{\bmath{J}},\sigma)} \over 
{\sigma - {\bmath{n}} \cdot {\bmath{\omega}} ({\bmath{J}})}} 
& {\bmath{n}} \cdot {{\upartial f_0} \over {\upartial {\bmath{J}}}}
 \exp ({\rm i} {\bmath{n}} \cdot {\bmath{w}} - {\rm i} \sigma t) \cr
&= \sum_{\bmath{n}} m_* \int {\rm d} \bmath{v}
\int {\rm d} {\bmath{\varpi}} \lambda ({\bmath{n}}, {\bmath{\varpi}}, \sigma)
\delta [{\bmath{J}} - {\bmath{J}}_R ({\bmath{n}} ,{\bmath{\varpi}},\sigma)]
\exp ({\rm i} {\bmath{n}} \cdot {\bmath{w}} - {\rm i} \sigma t) . \cr} \eqno\stepeq
$$
Equation (12) is a condition of self-consistency to be satisfied by perturbations of the system,
inasmuch as the density $\rho_1 ({\bmath{x}},t)$ must be the source of the gravitational
potential $V_1 ({\bmath{x}},t)$.  In other words, the characteristic value problem
governing the normal modes of the system now reduces to equations (3) and (12).  We can solve
the characteristic value problem with the aid of a modified version of the matrix
method of Kalnajs (1977).

Let $\rho_1 ({\bmath{x}},{\bmath{\beta}})$ and  $V_1 ({\bmath{x}},{\bmath{\beta}})$
be members of a biorthonormal set of densities and potentials, in the sense of
Clutton-Brock (1972) and Kalnajs (1976), where the constants ${\bmath{\beta}}$
are sets of parameters which label the members of that basis set, and each density
$\rho_1 ({\bmath{x}},{\bmath{\beta}})$ is the source of the corresponding potential
$V_1 ({\bmath{x}},{\bmath{\beta}})$.  A biorthonormal set of densities and potentials
satisfies the conditions
$$
\int_{\rm V} V_1^* ({\bmath{x}},{\bmath{\alpha}}) \rho_1 ({\bmath{x}},{\bmath{\beta}})
{\rm d} {\bmath{x}} = - \delta ( {\bmath{\alpha}} , {\bmath{\beta}} ) , \eqno\stepeq
$$
where the integration extends over the volume $ {\rm V} $ of the configuration,
and $\delta ( {\bmath{\alpha}} , {\bmath{\beta}} )$ is the Kronecker delta.  The minus
sign on the right-hand side of equation (13) is required by the physical interpretation
of the integral on the left-hand side as the gravitational potential energy of the
mass distribution described by the density $\rho_1 ({\bmath{x}},{\bmath{\beta}})$
in the gravitational potential $V_1^* ({\bmath{x}},{\bmath{\alpha}})$.  When
$\bmath{\beta} = \bmath{\alpha}$, the integral represents the self-energy of the
perturbation, which is negative.

The matrix formulation of the characteristic value problem is based on
representations of the perturbations of the density and the potential
of the forms
$$
\rho_1 ({\bmath{x}},t) = \sum_{\bmath{\beta}} \mu ({\bmath{\beta}},\sigma)
  \rho_1 ({\bmath{x}},{\bmath{\beta}}) \exp (- {\rm i} \sigma t)
\quad \hbox{ and } \quad
V_1 ({\bmath{x}},t) = \sum_{\bmath{\beta}} \mu ({\bmath{\beta}},\sigma)
  V_1 ({\bmath{x}},{\bmath{\beta}}) \exp (- {\rm i} \sigma t) , \eqno\stepeq
$$
respectively, where the coefficients $\mu ({\bmath{\beta}},\sigma)$ are constants
to be determined.  We shall need representations of the potentials
$V_1 ({\bmath{x}},{\bmath{\beta}})$ as trigonometric series in the angle variables
of the form
$$
V_1 ({\bmath{x}},{\bmath{\beta}}) = V_1 ({\bmath{w}},{\bmath{J}},{\bmath{\beta}}) 
= \sum_{\bmath{n}} V_1 ({\bmath{n}},{\bmath{J}},{\bmath{\beta}})
\exp ({\rm i} {\bmath{n}} \cdot {\bmath{w}}) . \eqno\stepeq
$$
It follows from the second of equations (14) that the coefficients in the trigonometric
series on the right-hand sides of equations (6) and (15) must be related in the manner
$$
V_1 (\bmath{n},\bmath{J},\sigma) = \sum_{\bmath{\beta}} \mu ({\bmath{\beta}},\sigma)
  V_1 (\bmath{n},\bmath{J},\bmath{\beta}) . \eqno\stepeq
$$

We obtain a matrix representation of the condition of self-consistency by multiplying each
term in equation (12) by $V_1^* (\bmath{x},\bmath{\alpha})$ and integrating the product over
the volume of the configuration space accessible to stars in the unperturbed system.  In the
reduction of the resulting equation, we transform integrals over the phase space from integrals
over $\bmath{x}$ and $\bmath{v}$ to integrals over $\bmath{w}$ and $\bmath{J}$.  Inasmuch as
this is a canonical transformation of coordinates and momenta, the Jacobian determinant of the
transformation is unity.  Making use of equations (13)-(16), and noting that integrals
of terms of the form $\exp [{\rm i} ({\bmath{n}} - {\bmath{n}^\prime}) \cdot {\bmath{w}}]$ over
the angle variables vanish unless ${\bmath{n}^\prime} = {\bmath{n}}$, where ${\bmath{n}}$
and ${\bmath{n}^\prime}$ are sets of integers, we reduce the resulting equation to
$$
\eqalign{
\sum_\bmath{\beta} \Big[ \delta (\bmath{\alpha},\bmath{\beta}) - \sum_\bmath{n}
(2 \upi)^3 m_* P \int {\rm d} \bmath{J}
{ { V_1^* (\bmath{n},\bmath{J},\bmath{\alpha})  V_1 (\bmath{n},\bmath{J},\bmath{\beta}) }
\over { \sigma - \bmath{n} \cdot \bmath{\omega} (\bmath{J}) } }
& \bmath{n} \cdot { { \upartial f_0 } \over { \upartial \bmath{J} } } \Big]
\mu (\bmath{\beta},\sigma) \cr
&= - \sum_\bmath{n} (2 \upi)^3 m_* \int {\rm d} \bmath{\varpi}
V_1^* [\bmath{n},\bmath{J}_R (\bmath{n},\bmath{\varpi},\sigma),\bmath{\alpha}]
\lambda (\bmath{n},\bmath{\varpi},\sigma) , \cr} \eqno\stepeq
$$
where we have suppressed a common factor $\exp ( - {\rm i} \sigma t)$.

\subsection{Normal modes}

Equation (17) represents a system of algebraic equations of the form
$$
\sum_{\bmath{\beta}} M (\sigma,{\bmath{\alpha}},{\bmath{\beta}}) \mu ({\bmath{\beta}},\sigma)
= \Lambda (\sigma,{\bmath{\alpha}}) \eqno\stepeq
$$
for the determination of the coefficients $\mu ({\bmath{\beta}},\sigma)$,
where
$$
M (\sigma,{\bmath{\alpha}} ,{\bmath{\beta}}) = \delta({\bmath{\alpha}},{\bmath{\beta}})
- \sum_{\bmath{n}} (2 \upi)^3 m_* 
P \int
{ {V_1^* ({\bmath{n}},{\bmath{J}},{\bmath{\alpha}}) V_1 ({\bmath{n}},{\bmath{J}},{\bmath{\beta}})} \over 
{\sigma - {\bmath{n}} \cdot {\bmath{\omega}} ({\bmath{J}})}}
{\bmath{n}} \cdot {{\upartial f_0} \over {\upartial {\bmath{J}}}} {\rm d} {\bmath{J}} \eqno\stepeq
$$
and
$$
\Lambda (\sigma,{\bmath{\alpha}}) = - \sum_{\bmath{n}} (2 \upi)^3 m_*
\int  V_1^* [{\bmath{n}},{\bmath{J}}_R ({\bmath{n}},{\bmath{\varpi}},\sigma),{\bmath{\alpha}}]
\lambda ({\bmath{n}},{\bmath{\varpi}},\sigma) {\rm d} {\bmath{\varpi}} . \eqno\stepeq
$$

\subsubsection{The continuous spectrum of modes}

As they stand, equations (18)-(20) provide a matrix representation of the
characteristic value problem governing modes with a continuous spectrum of real
frequencies.  For an assigned value of $\sigma$ in the continuous spectrum,
equations (18) are an inhomogeneous system of linear equations governing the
coefficients $\mu ({\bmath{\beta}},\sigma)$.  In general, those equations admit
of a solution only if ${\rm det} |M (\sigma,{\bmath{\alpha}},{\bmath{\beta}})| \neq 0$.

Without loss of generality, we can normalize the perturbation in such a way
that
$$
\int_{\rm V} V_1^* ({\bmath{x}},t) \rho_1 ({\bmath{x}},t) {\rm d} {\bmath{x}} = - 1 ,
\eqno\stepeq
$$
inasmuch as the gravitational potential energy of the perturbation must be negative.
In virtue of equations (13) and (14), this normalization reduces to
$$
\sum_{\bmath{\beta}} |\mu ({\bmath{\beta}}, \sigma)|^2 = 1. \eqno\stepeq
$$
For the assigned value of $\sigma$, equation (22) imposes one constraint on the
quantities $\Lambda (\sigma,{\bmath{\alpha}})$ and, hence, on the functions
$\lambda ({\bmath{n}},{\bmath{\varpi}},\sigma)$ (see equations [20]).  Inasmuch as the functions
$\lambda ({\bmath{n}},{\bmath{\varpi}},\sigma)$ are arbitrary, apart from that
constraint, there is no characteristic equation for the determination of the
value of $\sigma$.  Moreover, the modes belonging to an assigned value of
$\sigma$ in the continuous spectrum are highly degenerate.

The physical meaning of the degeneracy of the modes in the continuous spectrum
is the following.  The functions $\lambda ({\bmath{n}},{\bmath{\varpi}},\sigma)$ determine
the perturbation $f_1 ({\bmath{x}},{\bmath{v}},t)$ on the surfaces in the action space
on which $\sigma - {\bmath{n}} \cdot {\bmath{\omega}}({\bmath{J}}) = 0$ and the unperturbed
stellar orbits are resonant.  The arbitrariness of the
functions $\lambda ({\bmath{n}},{\bmath{\varpi}},\sigma)$ represents a freedom to
populate the resonant orbits in different ways in order to maintain
the selfconsistency of the perturbation.

One particular construction of modes of the van Kampen type can be achieved by choosing
$\lambda ({\bmath{n}},{\bmath{\varpi}},\sigma)$ in the manner
$$
\lambda ({\bmath{n}},{\bmath{\varpi}},\sigma)
= \delta(\bmath{n},\bmath{n}_0) \delta(\bmath{\varpi} - \bmath{\varpi}_0)
\lambda ({\bmath{n}}_0,{\bmath{\varpi}}_0,\sigma) , \eqno\stepeq
$$
where $\bmath{n}_0$ and $\bmath{\varpi}_0$ are fixed constants, $\delta(\bmath{n},\bmath{n}_0)$
denotes a Kronecker delta, and $\delta(\bmath{\varpi} - \bmath{\varpi}_0)$ denotes Dirac{\rq}s delta
function.  With this choice, we populate only the resonance $\bmath{n} = \bmath{n}_0$ with stars,
and, on the resonant surface $\sigma - \bmath{n}_0 \cdot \bmath{\omega} (\bmath{J}) = 0$ in the
action space, we include only stars whose actions have the values
$\bmath{J} = \bmath{J}_R (\bmath{n}_0,\bmath{\varpi}_0,\sigma)$.  In this case, equation (20)
reduces to
$$
\Lambda (\sigma,{\bmath{\alpha}}) = - (2 \upi)^3 m_*
V_1^* [{\bmath{n}}_0,{\bmath{J}}_R ({\bmath{n}}_0,{\bmath{\varpi}}_0,\sigma),{\bmath{\alpha}}]
\lambda ({\bmath{n}}_0,{\bmath{\varpi}}_0,\sigma) . \eqno\stepeq
$$
Therefore, all of the coefficients $\mu (\bmath{\beta},\sigma)$ in the solution of equation (18)
are proportional to $\lambda ({\bmath{n}}_0,{\bmath{\varpi}}_0,\sigma)$.  It follows that
the normalization of the coefficients specified by equation (22) determines the value of
$|\lambda ({\bmath{n}}_0,{\bmath{\varpi}}_0,\sigma)|^2$.

In Section 3, we shall find that nonlinear effects impose a constraint on the
frequencies of the modes of the van Kampen type.  Specifically, we shall find that the spectrum
of frequencies is bounded by the frequency of a mode of the \lq{Vlasov type.}\rq

\subsubsection{Modes of the Vlasov type}

At exceptional frequencies in the continuous spectrum, it can happen that
${\rm det} |M (\sigma,{\bmath{\alpha}},{\bmath{\beta}})| = 0$.  This is essentially a
characteristic equation for the determination of such exceptional values of $\sigma$.
Under these conditions, a
solution for the coefficients $\mu (\bmath{\beta},\sigma)$ exists if and only if
$\Lambda (\sigma,\bmath{\alpha}) = 0$ in equation (18).  In other words the quantities
$\lambda (\bmath{n},\bmath{\varpi},\sigma)$ must all vanish.  Nevertheless, the integrals that
appear in the definition of $M (\sigma,{\bmath{\alpha}},{\bmath{\beta}})$ in equation (19) must be
interpreted as Cauchy principal values.  In this case, the characteristic equation
${\rm det} |M (\sigma,{\bmath{\alpha}},{\bmath{\beta}})| = 0$ is the counterpart of the dispersion
relation proposed by Vlasov (1945) for plasma oscillations.  This is a mode in the continuous
spectrum in which stars on resonant orbits do not contribute to the self-consistency of
the perturbation.  The interpretation of Vlasov{\rq}s
dispersion relation in terms of an absence of particles trapped in a plasma wave has been discussed
by Bohm and Gross (1949) and Jackson (1960).  The interpretation of modes of the Vlasov
type is illuminated by the discussion of their slightly nonlinear counterparts at the
end of Section 3 below.

\subsubsection{The discrete spectrum of modes}

In the cases of modes with complex frequencies and of certain modes with real
frequencies, $\sigma - {\bmath{n}} \cdot {\bmath{\omega}}({\bmath{J}}) \neq 0$ for all values
of ${\bmath{n}}$ and all values ${\bmath{J}}$ accessible to stars in the unperturbed system.
In such cases, there are no terms in the sums over integrals involving the functions
$\lambda ({\bmath{n}},{\bmath{\varpi}},\sigma)$ on the right-hand sides of equations (12), (17),
and (20).  In other words, the quantities $\Lambda (\sigma,{\bmath{\alpha}})$ vanish identically,
and equations (18) reduce to
$$
\sum_{\bmath{\beta}} M (\sigma,{\bmath{\alpha}},{\bmath{\beta}}) \mu ({\bmath{\beta}},\sigma)
= 0 . \eqno\stepeq
$$
In this case, it is not necessary to interpret the integrals on the right-hand side of equation
(19) as Cauchy principal values.
Solutions of equations (25) for the coefficients $\mu ({\bmath{\beta}},\sigma)$ exist if
and only if ${\rm det} |M (\sigma,{\bmath{\alpha}},{\bmath{\beta}})| = 0$.  This condition
provides a characteristic equation for the determination of the frequencies $\sigma$
of the discrete modes.

For growing modes, the present version of the matrix method is equivalent to the version
of Kalnajs, because the prescription implied by equation (11) for the integration of quantities
involving the perturbation of the distribution function over the phase space coincides with
Landau{\rq}s prescription.  Accordingly, the two versions of the matrix method give the same
solutions for modes of instability.

\section{Slightly nonlinear oscillations of a galaxy}

In the theory of modes of the van Kampen type described in the preceding section, we have attributed
singularities in the perturbation of the distribution function to contributions of
stars on resonant orbits.  In the case of plasma oscillations, such an interpretation of the
singularities in the van Kampen modes has been verified by Bernstein, et al. (1957)
in their investigation of an exact model of nonlinear plasma waves (see also Stix 1992).

In this section, we investigate stationary oscillations of galaxies in cases in which the
amplitudes of the oscillations are small but finite.  As in the investigation of Bernstein, et al. (1957),
we find that such oscillations can be represented in terms of series in half-integral powers
of the amplitude of the perturbation of the gravitational potential.  We solve the governing equations
through terms of the first order in that amplitude.  The solution requires an
explicit treatment of the contributions of resonant orbits to the density of stars in the
six-dimensional phase space.  This treatment of stationary oscillations confirms and clarifies
the interpretation described above of the singularities in the modes of the van Kampen type.

\subsection{Formulation of the problem and reduction in terms of action-angle variables}

Consider an oscillating galaxy in which the density and the gravitational potential are of the forms
$$
\rho (\bmath{x},t) = \rho_0 (\bmath{x})
+ {1 \over 2} \epsilon \rho_1 (\bmath{x}) \exp (- {\rm i} \sigma t)
+ {1 \over 2} \epsilon \rho_1^* (\bmath{x}) \exp ( {\rm i} \sigma t) \eqno\stepeq
$$
and
$$
V (\bmath{x},t) = V_0 (\bmath{x})
+ {1 \over 2} \epsilon V_1 (\bmath{x}) \exp (- {\rm i} \sigma t)
+ {1 \over 2} \epsilon V_1^* (\bmath{x}) \exp ( {\rm i} \sigma t), \eqno\stepeq
$$
respectively, where $\sigma$ is a constant, real frequency and the small, dimensionless parameter
$\epsilon$ is a positive, real number.
We require that the oscillation be self-consistent in the sense that the density
$\rho (\bmath{x},t)$ is the source of the potential $V (\bmath{x},t)$.
Evidently, the density $\rho_0 (\bmath{x})$ must be the source of the gravitational potential
$V_0 (\bmath{x})$.  We shall find that $\rho_0 (\bmath{x})$ and $V_0 (\bmath{x})$ characterize an
unperturbed state of the galaxy.  Let $f_0 (\bmath{x},\bmath{v})$ denote the distribution function of
the galaxy in that unperturbed state.  In equations (26) and (27), we represent $\rho (\bmath{x},t)$
and $V (\bmath{x},t)$, respectively, as real quantities, expressed, for the sake of convenience in
what follows, in terms of a complex density $\rho_1 (\bmath{x})$ and a complex potential
$V_1 (\bmath{x})$, respectively.  The requirement of self-consistency also implies that
the density $\rho_1 (\bmath{x})$ must be the source of the potential $V_1 (\bmath{x})$.

The goal in what follows is to construct the distribution function $f(\bmath{x},\bmath{v},t)$ of
the galaxy in the state of oscillation described by equations (26) and (27).  In order to
accomplish this, we investigate the motions of stars in the gravitational potential described
by equation (27), we construct a set of integrals of the motion, and, with the aid of the theorem
of Jeans, we express $f(\bmath{x},\bmath{v},t)$ as a function of those integrals.  We carry out
this program under conditions in which the oscillatory perturbation of the system is small but
finite, and, in the construction of the distribution function, we evaluate all quantities involved
consistently through terms of the first order in the parameter $\epsilon$.  We have written the
expressions for $\rho (\bmath{x},t)$ and $V (\bmath{x},t)$ as we have in terms of $\epsilon$
in order to facilitate the ordering of terms in the series that arise in this process.

As in Section 2, we consider a system in which the motion of a star in the unperturbed potential
$V_0 (\bmath{x})$ is separable, and we introduce the actions $\bmath{J}$ and angles $\bmath{w}$
of the unperturbed motion as canonical momenta and coordinates, respectively.  The Hamiltonian
is expressible as a function $H = H_0 (\bmath{J})$ of the actions alone.  It follows from the
canonical equations of motion in that case that the actions $\bmath{J}$ are integrals of the
unperturbed motion of a star and the frequencies of the unperturbed motion are given by the
relations $\bmath{\omega} (\bmath{J}) = \upartial H_0 / \upartial \bmath{J}$.

We now express the function $V_1 (\bmath{x})$ as a function of the action-angle variables
and particularly as a trigonometric series in the angle variables.  Thus, we write
$$
V_1 ({\bmath{x}}) = V_1 ({\bmath{w}},{\bmath{J}}) 
= \sum_{\bmath{n}} V_1 ({\bmath{n}},{\bmath{J}})
\exp ({\rm i} {\bmath{n}} \cdot {\bmath{w}}) , \eqno\stepeq
$$
where the sum extends over sets of integers $\bmath{n} = (n_1,n_2,n_3)$.
The motion of a star in the potential $V(\bmath{x},t)$ is governed by the Hamiltonian
$$
\eqalign{ H(\bmath{w},\bmath{J},t) &= H_0 (\bmath{J})
+ {1 \over 2} \epsilon V_1 ({\bmath{w}},{\bmath{J}})\exp (- {\rm i} \sigma t)
+ {1 \over 2} \epsilon V_1^* ({\bmath{w}},{\bmath{J}}) \exp ( {\rm i} \sigma t) \cr
&= H_0 (\bmath{J})
+ {1 \over 2} \epsilon \sum_{\bmath{n}} [ V_1 ({\bmath{n}},{\bmath{J}})
\exp ({\rm i} {\bmath{n}} \cdot {\bmath{w}} - {\rm i} \sigma t)
+ V_1^* ({\bmath{n}},{\bmath{J}})
\exp ( - {\rm i} {\bmath{n}} \cdot {\bmath{w}} + {\rm i} \sigma t) ] . \cr } \eqno\stepeq
$$

\subsection{Construction of the integrals of the motion for a resonant orbit}

Consider now the motion of a star in the neighborhood of the resonance at which
$\sigma - \bmath{m} \cdot \bmath{\omega} (\bmath{J}) = 0$, where
$\bmath{m} = (m_1,m_2,m_3)$ is a particular set of three integers.  We can construct the
integrals of the motion through order $\epsilon$ in this case with the aid of
two successive canonical transformations.  This procedure is essentially von Zeipel{\rq}s
method for the construction of a canonical perturbation theory (von Zeipel 1916,
Contopoulos 1975, Goldstein 1980).

The first of these transformations replaces the actions $\bmath{J}$ and angles
$\bmath{w}$ with new momenta $\bmath{I} = (I_1,I_2,I_3)$ and new coordinates
$\bmath{\psi}= (\psi_1,\psi_2,\psi_3)$ with the aid of the generating function
$$
F(\bmath{I},\bmath{w},t) = \bmath{I} \cdot \bmath{w}
- {{\rm i} \over 2} \epsilon \sum_{\bmath{n} \ne \bmath{m}} \Big[ { { V_1 ({\bmath{n}},{\bmath{I}})
\exp ({\rm i} {\bmath{n}} \cdot {\bmath{w}} - {\rm i} \sigma t) } \over
{ \sigma - \bmath{n} \cdot \bmath{\omega} (\bmath{I}) } }
- { V_1^* ({\bmath{n}},{\bmath{I}})
\exp ( - {\rm i} {\bmath{n}} \cdot {\bmath{w}} + {\rm i} \sigma t) \over
{ \sigma - \bmath{n} \cdot \bmath{\omega} (\bmath{I}) } } \Big] , \eqno\stepeq
$$
a function of the new momenta and the old coordinates.  The transformation equations are
$$
\bmath{J} = {{\upartial F} \over {\upartial \bmath{w}}} = \bmath{I}
+ {1 \over 2} \epsilon \sum_{\bmath{n} \ne \bmath{m}} \Big[ { { \bmath{n} V_1 ({\bmath{n}},{\bmath{I}})
\exp ({\rm i} {\bmath{n}} \cdot {\bmath{w}} - {\rm i} \sigma t) } \over
{ \sigma - \bmath{n} \cdot \bmath{\omega} (\bmath{I}) } }
+ { \bmath{n} V_1^* ({\bmath{n}},{\bmath{I}})
\exp ( - {\rm i} {\bmath{n}} \cdot {\bmath{w}} + {\rm i} \sigma t) \over
{ \sigma - \bmath{n} \cdot \bmath{\omega} (\bmath{I}) } } \Big] \eqno\stepeq
$$
and
$$
\bmath{\psi} = {{\upartial F} \over {\upartial \bmath{I}}} = \bmath{w}
- {{\rm i} \over 2} \epsilon \sum_{\bmath{n} \ne \bmath{m}} \Big\{ 
 \Big[ { \upartial \over { \upartial \bmath{I} } }
\Big({{  V_1 ({\bmath{n}},{\bmath{I}}) } \over
{ \sigma - \bmath{n} \cdot \bmath{\omega} (\bmath{I}) } } \Big) \Big]
\exp ({\rm i} {\bmath{n}} \cdot {\bmath{w}} - {\rm i} \sigma t)
- \Big[ { \upartial \over { \upartial \bmath{I} } }
\Big({ { V_1^* ({\bmath{n}},{\bmath{I}}) } \over
{ \sigma - \bmath{n} \cdot \bmath{\omega} (\bmath{I}) } } \Big) \Big]
\exp ( - {\rm i} {\bmath{n}} \cdot {\bmath{w}} + {\rm i} \sigma t) \Big\} , \eqno\stepeq
$$
and the new Hamiltonian is
$$
K(\bmath{\psi},\bmath{I},t) = H(\bmath{w},\bmath{J},t) + { {\upartial F} \over {\upartial t}}
= H_0 (I)  +   {1 \over 2} \epsilon V_1 (\bmath{m},\bmath{I})
\exp ( {\rm i} {\bmath{m}} \cdot {\bmath{\psi}} - {\rm i} \sigma t)
+   {1 \over 2} \epsilon V_1^* (\bmath{m},\bmath{I})
\exp ( - {\rm i} {\bmath{m}} \cdot {\bmath{\psi}} + {\rm i} \sigma t)
+ O(\epsilon^2) \eqno\stepeq
$$
(Goldstein 1980).  In equation (33), we have eliminated the old angles $\bmath{w}$ in favor of
the new angles $\bmath{\psi}$ with the aid of equation (32).  This canonical transformation
removes the nonresonant terms of order $\epsilon$ from the perturbed Hamiltonian.

The second canonical transformation introduces final momenta $\bmath{G} = (G_R, G_2,G_1)$ and 
final coordinates $\bmath{\theta} = (\theta_R,\theta_2,\theta_3)$, and it is designed to make the quantity
${\bmath{m}} \cdot {\bmath{\psi}} - \sigma t$ a canonical coordinate.  The generating function is
$$
F(\bmath{G},\bmath{\psi},t) = G_R(\bmath{m} \cdot \bmath{\psi} - \sigma t) + G_2 \psi_2 + G_3 \psi_3 ,
\eqno\stepeq 
$$
again a function of the new momenta and the old coordinates.  The transformation equations are
$$
\theta_R = { {\upartial F } \over {\upartial G_R } } = \bmath{m} \cdot \bmath{\psi} - \sigma t ,
\quad \theta_2 = { {\upartial F } \over {\upartial G_2 } } = \psi_2 , \quad \hbox{ and } \quad
\theta_3 = { {\upartial F } \over {\upartial G_3 } } = \psi_3  \eqno\stepeq
$$
and
$$
I_1 = { {\upartial F} \over {\upartial \psi_1} } = m_1 G_R, \quad
I_2 = { {\upartial F} \over {\upartial \psi_2} } = m_2 G_R + G_2, \quad \hbox{ and } \quad
I_3 = { {\upartial F} \over {\upartial \psi_3} } = m_3 G_R + G_3 , \eqno\stepeq
$$
and the final Hamiltonian is
$$
\eqalign{ H_R (\theta_R,G_R ,\bmath{G}_\perp)
& = K(\bmath{\psi},\bmath{I},t) + {{\upartial F} \over {\upartial t}} \cr
= H_0 (\bmath{m} G_R & + \bmath{G}_\perp)  - \sigma G_R
+ {1 \over 2} \epsilon V_1 (\bmath{m},\bmath{m}G_R + \bmath{G}_\perp) \exp ( {\rm i} \theta_R)
+ {1 \over 2} \epsilon V_1^* (\bmath{m},\bmath{m}G_R + \bmath{G}_\perp) \exp ( - {\rm i} \theta_R)
+ O(\epsilon^2) \cr } \eqno\stepeq
$$
(Goldstein 1980), where we have defined $\bmath{G}_\perp = (0,G_2,G_3)$ and we have abbreviated
equations (36) to $\bmath{I} = \bmath{m}G_R + \bmath{G}_\perp$.

We observe that the new momenta $G_2$ and $G_3$ are integrals of the motion, inasmuch as the
Hamiltonian $H_R (\theta_R,G_R ,\bmath{G}_\perp)$ does not depend on the coordinates $\theta_2$
and $\theta_3$.  Thus $H_R (\theta_R,G_R ,\bmath{G}_\perp)$ is effectively a Hamiltonian governing
a motion in one degree of freedom in the canonical variables $(\theta_R,G_R)$.  The Hamiltonian
$H_R (\theta_R,G_R ,\bmath{G}_\perp)$ itself is a third integral of the motion inasmuch as it
does not depend explicitly on the time.

\subsection{Reduction of the perturbed Hamiltonian to a pendulum model}

Let $G_0$ be the \lq{resonance value}\rq\/ of $G_R$ satisfying the condition
$$
\bmath{m} \cdot \bmath{\omega} (\bmath{m}G_0 + \bmath{G}_\perp) - \sigma = 0 \eqno\stepeq
$$
for assigned values of the components $G_2$ and $G_3$ of $\bmath{G}_\perp$.  We expect, and
we shall find, that $G_R - G_0 = O(\epsilon^{1/2})$ for stars on resonant orbits.
In virtue of equation (38), an expansion of
the Hamiltonian $H_R (\theta_R,G_R ,\bmath{G}_\perp)$ in powers of $G_R - G_0$ accordingly yields,
in place of equation (37),
$$
\eqalign{ H_R (\theta_R,G_R ,\bmath{G}_\perp) = &H_0 (\bmath{m} G_0 + \bmath{G}_\perp) - \sigma G_0
+ {1 \over 2} \Big\{ { \upartial \over { \upartial G_0 } } 
[ \bmath{m} \cdot \bmath{\omega} (\bmath{m}G_0 + \bmath{G}_\perp) ] \Big\} (G_R - G_0)^2 \cr
&+ {1 \over 2} \epsilon V_1 (\bmath{m},\bmath{m}G_0 + \bmath{G}_\perp) \exp ( {\rm i} \theta_R)
+ {1 \over 2} \epsilon V_1^* (\bmath{m},\bmath{m}G_0 + \bmath{G}_\perp) \exp ( - {\rm i} \theta_R)
+ O(\epsilon^{3/2}) . \cr } \eqno\stepeq
$$

Let
$$
\eta = {\rm sign} \Big\{ { \upartial \over {\upartial G_0}}
[\bmath{m} \cdot \bmath{\omega} (\bmath{m}G_0 + \bmath{G}_\perp)] \Big\} \quad \hbox{ and } \quad
\mu = \Big| { \upartial \over {\upartial G_0}}
[\bmath{m} \cdot \bmath{\omega} (\bmath{m}G_0 + \bmath{G}_\perp)] \Big| , \eqno\stepeq
$$
and write the function $V_1 (\bmath{m},\bmath{m}G_0 + \bmath{G}_\perp)$ in terms of its
amplitude and phase in the manner
$$
V_1 (\bmath{m},\bmath{m}G_0 + \bmath{G}_\perp) = |V_1 (\bmath{m},\bmath{m}G_0 + \bmath{G}_\perp)|
\exp ({\rm i} \phi)  .  \eqno\stepeq
$$
Then, equation (39) reduces to
$$
H_R (\theta_R,G_R ,\bmath{G}_\perp) = H_0 (\bmath{m} G_0 + \bmath{G}_\perp) - \sigma G_0
+ \eta \Big[ {1 \over 2} \mu (G_R - G_0)^2 +
\eta \epsilon |V_1 (\bmath{m},\bmath{m}G_0 + \bmath{G}_\perp)| \cos (\theta_R + \phi) \Big]
+ O(\epsilon^{3/2}) . \eqno\stepeq
$$

This Hamiltonian has a stable fixed point at $(G_R,\theta_R) = (G_0,\theta_0)$, where the
value of $\theta_0$ is such that $\cos (\theta_0 + \phi) = - \eta$.  Inasmuch as the value
of $\eta$ is either $1$ or $-1$ (see equations [40]), it follows that an alternative
form of equation (42) is
$$
H_R (\theta_R,G_R ,\bmath{G}_\perp) = H_0 (\bmath{m} G_0 + \bmath{G}_\perp) - \sigma G_0
+ \eta \Big[ {1 \over 2} \mu (G_R - G_0)^2 -
\epsilon |V_1 (\bmath{m},\bmath{m}G_0 + \bmath{G}_\perp)| \cos (\theta_R - \theta_0) \Big]
+ O(\epsilon^{3/2}) . \eqno\stepeq
$$
This is essentially the Hamiltonian of a physical pendulum in the canonical momentum
$\Delta G = G_R - G_0$ and the conjugate coordinate $\Delta \theta = \theta_R - \theta_0$.
The separatrix of the system, for an assigned value of $\bmath{G}_\perp$, satisfies the equation 
$$
{1 \over 2} \mu (G_R - G_0)^2 -
\epsilon |V_1 (\bmath{m},\bmath{m}G_0 + \bmath{G}_\perp)| \cos (\theta_R - \theta_0)
= \epsilon |V_1 (\bmath{m},\bmath{m}G_0 + \bmath{G}_\perp)| , \eqno\stepeq
$$
where we are now ignoring terms of orders $\epsilon^{3/2}$ and higher in the Hamiltonian.
In other words, the region of the phase space in which the angle $\theta_R$ librates about the
value $\theta_0$ and the motion is resonant is defined by the inequality $G_- \le G_R \le G_+$, where
$$
G_\pm = G_0 \pm \epsilon^{1/2} \Big[ { 2 \over \mu }
|V_1 (\bmath{m},\bmath{m}G_0 + \bmath{G}_\perp)| \Big]^{1/2}
[ 1 + \cos(\theta_R - \theta_0)]^{1/2} . \eqno\stepeq
$$
This definition of the resonance region confirms the expectation described above that
$G_R - G_0 = O(\epsilon^{1/2})$ for stars in the resonance region.  In other words, the size
of the resonance region is of order $\epsilon^{1/2}$.  Outside the resonance region, the
angle $\theta_R$ precesses, and the motion is essentially nonresonant.

The resonance region described by equation (43) for the Hamiltonian and equation (44) for the
separatrix consists of one or more islands in a closed chain in the phase space.  The island or
chain of islands circulates in the phase space.  The guiding centre for the circulation of each
island is a periodic orbit whose motion is given by
$\bmath{J} = \bmath{m} G_R + \bmath{G}_\perp + O(\epsilon)$ and
$\bmath{m} \cdot \bmath{w} - \sigma t + O(\epsilon) = \theta_0$.

\subsection{A matrix representation of a self-consistent oscillation of a galaxy}

We are now ready to make use of the theorem of Jeans in order to construct the distribution
function $f(\bmath{x},\bmath{v},t)$ underlying the density distribution and the gravitational
potential described by equations (26) and (27), respectively.  We shall construct the distribution
function in the non-resonance and resonance regions of the phase space separately.  Moreover,
we shall consider a system in which only a single resonance, defined by the set of integers
$\bmath{m} = (m_1,m_2,m_3)$, is populated with stars.  Finally, we shall develop
the distribution function as a series in powers of $\epsilon$, working consistently to an
order of approximation such that the fundamental relation
$$
m_* \int f(\bmath{x},\bmath{v},t) {\rm d} \bmath{v} = \rho(\bmath{x},t)  \eqno\stepeq
$$
is satisfied through terms of order $\epsilon$.

In this connection, we recall that we have identified the functions $\rho_0 (\bmath{x})$ and
$V_0 (\bmath{x})$ in equations (26) and (27) as the density and gravitational potential,
respectively, that characterize an unperturbed state of the galaxy.  According to the
theorem of Jeans, the distribution function $f_0 (\bmath{x},\bmath{v})$ for the galaxy in
that state can be expressed as $f_0 (\bmath{J})$, a function of the unperturbed action
integrals.  The self-consistency of the equilibrium of the galaxy in that state requires that
$$
m_* \int f_0 (\bmath{x},\bmath{v}) {\rm d} \bmath{v} = m_* \int f_0 (\bmath{J}) {\rm d} \bmath{v}
= \rho_0 (\bmath{x}). \eqno\stepeq
$$

\subsubsection{Non-resonant stars}

We first construct the distribution function for non-resonant stars, that is, for stars outside
the region of the phase space defined by the inequality $G_- \le G_R \le G_+$, where $G_\pm$
is defined by equation (45).  For the construction of the integrals of the motion required
for this purpose, we modify slightly the canonical transformation  described by equations (30)-(32).
In the trigonometric series in those equations, we now include the terms $\bmath{n} = \bmath{m}$
which were previously omitted.  In place of equation (33), the equation for the new Hamiltonian is now
$K(\bmath{\psi},\bmath{I},t) = H_0(\bmath{I}) + O(\epsilon^2)$.  In this case, the canonical
equations of motion imply that the new momenta $\bmath{I}$ are integrals of the motion through
terms of order $\epsilon$.  According to the theorem of Jeans, the distribution function (for the
non-resonant stars) can be expressed as a function of the integrals $\bmath{I}$.  Inasmuch as
the unperturbed distribution function of the system is of the form
$f_0 (\bmath{x},\bmath{v}) = f_0 (\bmath{J})$ described above, we write the perturbed distribution
function for the non-resonant stars as
$$
f_{NR} (\bmath{x},\bmath{v},t) = f_0 (\bmath{I})
= f_0 (\bmath{J})
- {1 \over 2} \epsilon \sum_{\bmath{n}} \Big[ { { V_1 ({\bmath{n}},{\bmath{J}})
\exp ({\rm i} {\bmath{n}} \cdot {\bmath{w}} - {\rm i} \sigma t) } \over
{ \sigma - \bmath{n} \cdot \bmath{\omega} (\bmath{J}) } }
+ { V_1^* ({\bmath{n}},{\bmath{J}})
\exp ( - {\rm i} {\bmath{n}} \cdot {\bmath{w}} + {\rm i} \sigma t) \over
{ \sigma - \bmath{n} \cdot \bmath{\omega} (\bmath{J}) } } \Big]
\bmath{n} \cdot {{\upartial f_0} \over {\upartial \bmath{J}}}
+ O(\epsilon^2) \eqno\stepeq
$$
(see eq. [31], but recall that we now include the terms $\bmath{n} = \bmath{m}$ in the
trigonometric series there).

The physical picture underlying this choice for the non-resonant distribution function is
that the non-resonant orbits are orbits which are not altered qualitatively by the perturbation.
In particular, the perturbed angle variables $\bmath{\psi}$ precess as do the unperturbed
angle variables $\bmath{w}$.  Thus, although the non-resonant orbits are perturbed, the
dependence of the distribution function on the integrals of the motion is not altered for the
non-resonant orbits.

The contribution of the non-resonant stars to the density of the system is
$$
\eqalign{ \rho_{NR} (\bmath{x}, &t) = m_* \int_{NR} f_0 (\bmath{I}) {\rm d} \bmath{v} \cr
= m_* &\int_{NR} f_0 (\bmath{J}) {\rm d} \bmath{v} - m_* P \int_{NR}
{1 \over 2} \epsilon \sum_{\bmath{n}} \Big[ { { V_1 ({\bmath{n}},{\bmath{J}})
\exp ({\rm i} {\bmath{n}} \cdot {\bmath{w}} - {\rm i} \sigma t) } \over
{ \sigma - \bmath{n} \cdot \bmath{\omega} (\bmath{J}) } }
+ { V_1^* ({\bmath{n}},{\bmath{J}})
\exp ( - {\rm i} {\bmath{n}} \cdot {\bmath{w}} + {\rm i} \sigma t) \over
{ \sigma - \bmath{n} \cdot \bmath{\omega} (\bmath{J}) } } \Big]
\bmath{n} \cdot {{\upartial f_0} \over {\upartial \bmath{J}}} {\rm d} \bmath{v}
+ O(\epsilon^2) , \cr } \eqno\stepeq
$$
where the integrals include only regions of the phase space in which the
stellar orbits are non-resonant, and $P$ signifies that integrals of functions containing
resonance denominators are to be interpreted as Cauchy principal values.  We reduce 
equation (49) to the more useful form
$$
\eqalign{ \rho_{NR} (\bmath{x}, t) = \rho_0 (\bmath{x})
&- m_* \int_{R} f_0 (\bmath{J}) {\rm d} \bmath{v} \cr
&- m_* P \int
{1 \over 2} \epsilon \sum_{\bmath{n}} \Big[ { { V_1 ({\bmath{n}},{\bmath{J}})
\exp ({\rm i} {\bmath{n}} \cdot {\bmath{w}} - {\rm i} \sigma t) } \over
{ \sigma - \bmath{n} \cdot \bmath{\omega} (\bmath{J}) } }
+ { V_1^* ({\bmath{n}},{\bmath{J}})
\exp ( - {\rm i} {\bmath{n}} \cdot {\bmath{w}} + {\rm i} \sigma t) \over
{ \sigma - \bmath{n} \cdot \bmath{\omega} (\bmath{J}) } } \Big]
\bmath{n} \cdot {{\upartial f_0} \over {\upartial \bmath{J}}} {\rm d} \bmath{v}
+ O(\epsilon^{3/2}) , \cr } \eqno\stepeq
$$
where we have rewritten the integral of $f_0 (\bmath{J})$ with the aid of equation (47) and
we have made use of the label $R$ in order to signify that an integral includes only the region
of the phase space in which the orbits are resonant.  In equation (50), we have also enlarged
the region of integration of the last integral in equation (49) to include the region of the
phase space containing resonant orbits.  Inasmuch as the size of the resonant region is of
order $\epsilon^{1/2}$, the result is to change the integral by a quantity of order $\epsilon^{3/2}$.
That is beyond the order of approximation required here.

The interpretation of integrals in equations (49) and (50) as Cauchy principal values now applies
at the resonance $\bmath{n} = \bmath{m}$ as well as at all other resonances.  The reduction of
equation (49) to equation (50) shows that, in general, the interpretation of the integrals as
Cauchy principal values neglects the finite sizes of the resonance regions in the phase space
and consequently neglects corrections of order $\epsilon^{3/2}$ to the density.

\subsubsection{Resonant stars}

We have found in Section 3.2 that the Hamiltonian $H_R (\theta_R,G_R,\bmath{G}_\perp)$ and
the components $G_2$ and $G_3$ of $\bmath{G}_\perp$ are integrals of the motion in the resonance
region of the phase space.  According to the theorem of Jeans, the distribution function
for the resonant stars can be written as
$$
f_R (\bmath{x},\bmath{v},t) = f_R (H_R,\bmath{G}_\perp) , \eqno\stepeq
$$
where the function $f_R (H_R,\bmath{G}_\perp)$ remains to the determined (see Section 3.5).

The contribution of the resonant stars to the density is
$$
\rho_R (\bmath{x},t) = m_* \int_R f_R (H_R,\bmath{G}_\perp) {\rm d} \bmath{v}.
\eqno\stepeq
$$

\subsubsection{A matrix representation of self-consistency}

The condition that the oscillation of the galaxy be self-consistent requires that the sum
of the densities $\rho_{NR} (\bmath{x}, t)$ and $\rho_R (\bmath{x},t)$ be equal to
the \lq{imposed}\rq\/ density $\rho (\bmath{x}, t)$.  Combining equations (26), (50), and
(52), canceling the term $\rho_0 (\bmath{x})$ on the two sides of the resulting equation, and
rearranging terms in what is left of the resulting equation, we obtain
$$
\eqalign{
m_* \int_R f_R (H_R,&\bmath{G}_\perp) {\rm d} \bmath{v}
- m_* \int_{R} f_0 (\bmath{J}) {\rm d} \bmath{v}
= {1 \over 2} \epsilon \rho_1 (\bmath{x}) \exp (- {\rm i} \sigma t)
+ {1 \over 2} \epsilon \rho_1^* (\bmath{x}) \exp ( {\rm i} \sigma t) \cr
+  &m_* P \int
{1 \over 2} \epsilon \sum_{\bmath{n}} \Big[ { { V_1 ({\bmath{n}},{\bmath{J}})
\exp ({\rm i} {\bmath{n}} \cdot {\bmath{w}} - {\rm i} \sigma t) } \over
{ \sigma - \bmath{n} \cdot \bmath{\omega} (\bmath{J}) } }
+ { V_1^* ({\bmath{n}},{\bmath{J}})
\exp ( - {\rm i} {\bmath{n}} \cdot {\bmath{w}} + {\rm i} \sigma t) \over
{ \sigma - \bmath{n} \cdot \bmath{\omega} (\bmath{J}) } } \Big]
\bmath{n} \cdot {{\upartial f_0} \over {\upartial \bmath{J}}} {\rm d} \bmath{v}
+ O(\epsilon^{3/2}) . \cr } \eqno\stepeq
$$
Equation (53), which expresses the condition of self-consistency for a slightly nonlinear
oscillation of the system, is to be compared (apart from terms of orders $\epsilon^{3/2}$
and higher) with equation (12), which expresses the condition
of self-consistency for an infinitesimal perturbation of the system.  In particular, the
right-hand side of equation (53) is essentially the real part of the left-hand side of
equation (12).  Thus, the essential difference between the two equations lies in the
different representations of the contributions of the resonant stars on the right-hand
side of equation (12) and on the left-hand side of equation (53).

Beginning with equation (53), we can now formulate the matrix method for the determination of
the functions $\rho_1 (\bmath{x})$, $V_1 (\bmath{x})$, and $f_R (H_R,\bmath{G}_\perp)$.
In this formulation, we represent  $\rho_1 (\bmath{x})$ and $V_1 (\bmath{x})$ in terms of
the biorthonormal set of densities and potentials introduced in Section 2.2.  In particular,
we now let equations (13)-(16) apply to
$\rho_1 (\bmath{x},t) = \rho_1 (\bmath{x}) \exp(-{\rm i} \sigma t)$
and $V_1 (\bmath{x},t) = V_1 (\bmath{x}) \exp(-{\rm i} \sigma t)$.

We next multiply each term in equation (53) by
$V_1^* (\bmath{x},\bmath{\alpha}) \exp({\rm i} \sigma t)$, integrate the result over
the configuration space, and then average the result over one period $2 \upi / \sigma$ of the
oscillation.  The reduction here of the terms appearing on the right-hand side of equation (53)
follows closely the similar reduction in Section 2.2 of terms appearing on the left-hand side
of equation (12) and leading to the terms on the left-hand side of equation (17).  In other words,
the immediate result of these operations on the terms in equation (53) is
$$
\eqalign{
{ \sigma \over { 2 \upi}} \int_0^{2 \upi / \sigma} {\rm d}t \exp({\rm i} \sigma t)
& m_* \int_R {\rm d} \bmath{x} {\rm d} \bmath{v} V_1^* (\bmath{x},\bmath{\alpha})
[ f_R (H_R,\bmath{G}_\perp) - f_0 (\bmath{J}) ] \cr
&= - {1 \over 2} \epsilon
\sum_\bmath{\beta} \Big[ \delta (\bmath{\alpha},\bmath{\beta}) - \sum_\bmath{n}
(2 \upi)^3 m_* P \int {\rm d} \bmath{J}
{ { V_1^* (\bmath{n},\bmath{J},\bmath{\alpha})  V_1 (\bmath{n},\bmath{J},\bmath{\beta}) }
\over { \sigma - \bmath{n} \cdot \bmath{\omega} (\bmath{J}) } }
 \bmath{n} \cdot { { \upartial f_0 } \over { \upartial \bmath{J} } } \Big]
\mu (\bmath{\beta},\sigma) + O(\epsilon^{3/2}) . \cr } \eqno\stepeq
$$

On the left-hand side of equation (54), we transform the integral over the resonance region
of the phase space to an integral over the action-angle variables $\bmath{J}$ and $\bmath{w}$
and we recall that the Jacobian determinant of the transformation is unity.
Writing $V_1^* (\bmath{x},\bmath{\alpha})$ as a trigonometric series in the angle variables
in accordance with equation (15), we obtain
$$
m_* \int_R {\rm d} \bmath{x} {\rm d} \bmath{v} V_1^* (\bmath{x},\bmath{\alpha})
[ f_R (H_R,\bmath{G}_\perp) - f_0 (\bmath{J}) ]
= \sum_\bmath{n} m_* \int_R {\rm d} \bmath{w} {\rm d} \bmath{J}
V_1^* (\bmath{n},\bmath{J},\bmath{\alpha}) \exp (- {\rm i} \bmath{n} \cdot \bmath{w})
[ f_R (H_R,\bmath{G}_\perp) - f_0 (\bmath{J}) ] . \eqno\stepeq
$$

We next transform the integral over the actions $\bmath{J}$ on the right-hand side of
equation (55) to an integral over the momenta $G_R$, $G_2$, and $G_3$.  For this purpose,
we make use of the transformation equations (31), (32), (35), and (36), and we observe
accordingly that
$$
\bmath{J} = \bmath{m}G_R + \bmath{G}_\perp + O(\epsilon)
= \bmath{m}G_0 + \bmath{G}_\perp + \bmath{m}(G_R - G_0) + O(\epsilon),
\quad \hbox{ and } \quad
{\rm d} \bmath{J} = m_1[1 + O(\epsilon)] {\rm d}G_R {\rm d}G_2 {\rm d}G_3 .
$$
In the transformed integral, the integration over $G_R$ is restricted to the resonance
region $G_- \le G_R \le G_+$, where $G_\pm$ is given by equation (45).  Inasmuch as
$J_1 = m_1 G_R + O(\epsilon)$, the limits of integration over $G_R$ depend on the
algebraic sign of $m_1$.  Thus, if $m_1 > 0$, then the upper and lower limits of integration
are $G_+$ and $G_-$, respectively.  However, if $m_1 < 0$, then the upper and lower limits
of integration are $G_-$ and $G_+$, respectively.  In order to include both cases in
equation (56) below, we replace $m_1 {\rm d} G_R$ with $| m_1 | {\rm d} G_R$ and consistently
adopt $G_+$ and $G_-$ as the upper and lower limits of integration, respectively.

With the further observation that
$$
f_0 (\bmath{J}) = f_0 (\bmath{m}G_0 + \bmath{G}_\perp) +
\Big[ {\upartial \over {\upartial G_0}}f_0 (\bmath{m}G_0 + \bmath{G}_\perp) \Big]
(G_R - G_0) + O(\epsilon)
$$
and
$$
V_1^* (\bmath{n},\bmath{J},\bmath{\alpha})
= V_1^* (\bmath{n},\bmath{m}G_0 + \bmath{G}_\perp,\bmath{\alpha})
+\Big[ {\upartial \over {\upartial G_0}}
V_1^* (\bmath{n},\bmath{m}G_0 + \bmath{G}_\perp,\bmath{\alpha}) \Big]
(G_R - G_0) + O(\epsilon) ,
$$
we finally reduce equation (55) to
$$
\eqalign{
m_* & \int_R {\rm d} \bmath{x} {\rm d} \bmath{v} V_1^* (\bmath{x},\bmath{\alpha})
[ f_R (H_R,\bmath{G}_\perp) - f_0 (\bmath{J}) ] \cr
= & \sum_\bmath{n} m_* \int {\rm d} \bmath{w}  \exp (- {\rm i} \bmath{n} \cdot \bmath{w})
\int {\rm d}G_2 {\rm d}G_3 V_1^* (\bmath{n},\bmath{m}G_0 + \bmath{G}_\perp,\bmath{\alpha})
| m_1 | \int_{G_-}^{G_+}
[ f_R (H_R,\bmath{G}_\perp) - f_0 (\bmath{m}G_0 + \bmath{G}_\perp) ] {\rm d}G_R
[1 + O(\epsilon)] . \cr } \eqno\stepeq
$$
Contributions of order $\epsilon^{1/2}$ vanish on the right-hand side of equation (56),
because they involve integrals over $G_R$ of odd functions of $(G_R - G_0)$ with limits
of integration which are symmetric with respect to the value $G_R = G_0$.

In the next subsection, we shall seek a solution for $f_R (H_R,\bmath{G}_\perp)$ such
that
$$
| m_1 | \int_{G_-}^{G_+}
[ f_R (H_R,\bmath{G}_\perp) - f_0 (\bmath{m}G_0 + \bmath{G}_\perp) ] {\rm d}G_R
= {1 \over 2} \epsilon \lambda(\bmath{m},\bmath{G}_\perp,G_0) \exp({\rm i} \theta_R)
+ {1 \over 2} \epsilon \lambda^* (\bmath{m},\bmath{G}_\perp,G_0) \exp( - {\rm i} \theta_R) ,
\eqno\stepeq
$$
where $\lambda(\bmath{m},\bmath{G}_\perp,G_0)$ is a function to be determined.  In the
remainder of this subsection, we shall complete the formulation of the matrix method.
Accordingly, we now combine equations (56) and (57) in order to rewrite the left-hand
side of equation (54).  Inasmuch as $\exp({\rm i} \theta_R) 
= \exp({\rm i} \bmath{m} \cdot \bmath{w} - {\rm i} \sigma t) [1 + O(\epsilon)]$
(see eqs. [32] and [35]), we bring equation (54) to the form
$$
\eqalign{
{1 \over 2} (2\upi)^3 \epsilon m_*
\int & {\rm d}G_2 {\rm d}G_3 V_1^* (\bmath{m},\bmath{m}G_0 + \bmath{G}_\perp,\bmath{\alpha})
 \lambda(\bmath{m},\bmath{G}_\perp,G_0) \cr
&= - {1 \over 2} \epsilon
\sum_\bmath{\beta} \Big[ \delta (\bmath{\alpha},\bmath{\beta}) - \sum_\bmath{n}
(2 \upi)^3 m_* P \int {\rm d} \bmath{J}
{ { V_1^* (\bmath{n},\bmath{J},\bmath{\alpha})  V_1 (\bmath{n},\bmath{J},\bmath{\beta}) }
\over { \sigma - \bmath{n} \cdot \bmath{\omega} (\bmath{J}) } }
 \bmath{n} \cdot { { \upartial f_0 } \over { \upartial \bmath{J} } } \Big]
\mu (\bmath{\beta},\sigma) + O(\epsilon^{3/2}) . \cr } \eqno\stepeq
$$

Equation (58), with the terms of orders $\epsilon^{3/2}$ and higher suppressed,
is an inhomogeneous, linear equation for the determination of the coefficients
$\mu (\bmath{\beta},\sigma)$.  To order $\epsilon$, equation (58) is essentially
the matrix representation of self-consistency obtained previously in equation (17) for modes
of the van Kampen type.  The most substantial difference between equations (17) and (58) is that
the inhomogeneous terms on the right-hand side of equation (17) represent a sum of contributions
of resonant stars, whereas the inhomogeneous term on the left-hand side of equation
(58) represents the contribution of a single resonance.  The remaining differences between
the two equations are entirely superficial.  The components $G_2$ and $G_3$ of $\bmath{G}_\perp$,
which label points on the surface
$\sigma - \bmath{m} \cdot \bmath{\omega}(\bmath{m}G_0 + \bmath{G}_\perp ) = 0$ in the action space,
are equivalent to the components $\varpi_1$ and $\varpi_2$ of $\bmath{\varpi}$, which label
points on the surfaces $\sigma - \bmath{n} \cdot \bmath{\omega}(\bmath{J}) = 0$.
Likewise, the quantity $\bmath{m}G_0 + \bmath{G}_\perp$ is equivalent to
the function $J_R (\bmath{n},\bmath{\varpi},\sigma)$.  The dependence of
$\lambda(\bmath{m},\bmath{G}_\perp,G_0)$ on $G_0$ is implicitly a dependence on
$\bmath{m}$, $\sigma$, and $\bmath{G}_\perp$ in virtue of equation (38).

Equation (58) is readily generalized in order to include contributions of a number of resonances
on the left-hand side.  For this purpose, the analysis in Sections 3.2, 3.3 and 3.4.2 would be
performed within each of the resonance regions that contain stars, and the results incorporated
in equations (53)-(58).  In each of the resonance regions included, the solution for the
perturbation of the distribution function would be determined as is described in Section 3.5.

The solution of equation (58) for the coefficients $\mu (\bmath{\beta},\sigma)$ can be obtained
and the constraint on $\lambda(\bmath{m},\bmath{G}_\perp,G_0)$ imposed along the lines described
in Sections 2.2 and 2.3 in the case of equation (17).  However, in the case of slightly nonlinear
oscillations, we shall find in the next subsection that the solution of equation (57) for
$f_R (H_R,\bmath{G}_\perp)$ imposes additional constraints on $\lambda(\bmath{m},\bmath{G}_\perp,G_0)$
which do not arise in the linear theory of modes of the van Kampen type.

\subsection{Construction of the distribution function for the resonant stars}

We shall now solve equation (57) for $f_R (H_R,\bmath{G}_\perp)$.  Let
$$
f_R (H_R,\bmath{G}_\perp) = f_0 (\bmath{m}G_0 + \bmath{G}_\perp)
+ F_R (E_R ;\bmath{m},\bmath{G}_\perp,G_0) , \eqno\stepeq
$$
where $F_R (E_R ;\bmath{m},\bmath{G}_\perp,G_0)$ is the component of
$f_R (H_R,\bmath{G}_\perp)$ which remains to be determined, we have defined an
{\lq}energy{\rq}
$$
\eqalign{ E_R &= \eta \{ H_R (\theta_R,G_R ,\bmath{G}_\perp)
- [H_0 (\bmath{m} G_0 + \bmath{G}_\perp) - \sigma G_0
+ \eta \epsilon |V_1 (\bmath{m},\bmath{m}G_0 + \bmath{G}_\perp)|] \} \cr
&= {1 \over 2} \mu (\Delta G)^2 -
\epsilon |V_1 (\bmath{m},\bmath{m}G_0 + \bmath{G}_\perp)| [ 1 + \cos (\Delta \theta)]
\cr } \eqno\stepeq
$$
(see eq.[43]), and we have introduced new canonical variables $\Delta G = G_R - G_0$ and
$\Delta \theta = \theta_R - \theta_0$.  Then we can rewrite equation (57) in the form
$$
| m_1 | \int_{G_- - G_0}^{G_+ -G_0}
 F_R (E_R ;\bmath{m},\bmath{G}_\perp,G_0) {\rm d}(\Delta G)
= \epsilon {\rm Re}[ \lambda(\bmath{m},\bmath{G}_\perp,G_0) \exp({\rm i} \theta_0)] \cos(\Delta \theta)
- \epsilon {\rm Im}[ \lambda(\bmath{m},\bmath{G}_\perp,G_0) \exp({\rm i} \theta_0)] \sin(\Delta \theta) .
\eqno\stepeq
$$
It follows from equations (45) and (60) that the integral on the left-hand side of equation (61)
must be a function of $\cos (\Delta \theta)$ and, hence, an even function of $\Delta \theta$.
Therefore, equation (61) can be satisfied only if
$$
{\rm Im}[ \lambda(\bmath{m},\bmath{G}_\perp,G_0) \exp({\rm i} \theta_0)] = 0.  \eqno\stepeq
$$
In other words, the quantity $\lambda(\bmath{m},\bmath{G}_\perp,G_0) \exp({\rm i} \theta_0)$
must be real, and equation (61) reduces to
$$
| m_1 | \int_{G_- - G_0}^{G_+ -G_0}
 F_R (E_R ;\bmath{m},\bmath{G}_\perp,G_0) {\rm d}(\Delta G)
= \epsilon \lambda(\bmath{m},\bmath{G}_\perp,G_0) \exp({\rm i} \theta_0) \cos(\Delta \theta) .
\eqno\stepeq
$$
Equation (63) can be transformed into an integral equation of the Abelian form.  Making use
of equation (60) in the inversion of equation (63), we obtain the solution
$$
F_R (E_R ;\bmath{m},\bmath{G}_\perp,G_0)
= - \Big({\mu \over 2}\Big)^{1/2}
{ { \lambda(\bmath{m},\bmath{G}_\perp,G_0) \exp({\rm i} \theta_0) } \over
{ | m_1 | \upi |V_1 (\bmath{m},\bmath{m}G_0 + \bmath{G}_\perp)|}}
[ \epsilon |V_1 (\bmath{m},\bmath{m}G_0 + \bmath{G}_\perp)| (-E_R)^{-1/2}
-2(-E_R)^{1/2}]  .  \eqno\stepeq
$$
Upon combining this result with equation (59), we obtain the distribution function for the
resonant stars in the form
$$
f_R (H_R,\bmath{G}_\perp)
= f_0 (\bmath{m}G_0 + \bmath{G}_\perp)
- \Big({\mu \over 2}\Big)^{1/2}
{ { \lambda(\bmath{m},\bmath{G}_\perp,G_0) \exp({\rm i} \theta_0) } \over
{ | m_1 | \upi |V_1 (\bmath{m},\bmath{m}G_0 + \bmath{G}_\perp)|}}
[ \epsilon |V_1 (\bmath{m},\bmath{m}G_0 + \bmath{G}_\perp)| (-E_R)^{-1/2}
-2(-E_R)^{1/2}]  .  \eqno\stepeq
$$

The term $f_0 (\bmath{m}G_0 + \bmath{G}_\perp)$ on the right-hand side of equation (65)
is of order unity, and it describes a uniform density of stars in the resonance region
of the phase space.  Moreover, the contribution of $f_0 (\bmath{m}G_0 + \bmath{G}_\perp)$
to the density of stars in the resonance region and the contribution of $f_0 (\bmath{J})$
to the density in the non-resonance region (see eq. [48]) combine in equation (46) to
give the unperturbed density $\rho_0 (\bmath{x})$ on the right-hand side of equation (26).

The second term on the right-hand side of equation (65)
is formally of order $\epsilon^{1/2}$, inasmuch as $E_R$ is of order $\epsilon$.  That
contribution to the distribution function in the resonance region and the contribution
of the terms of order $\epsilon$ on the right-hand side of equation (48) to the distribution
function in the non-resonance region combine in equation (46) to give the sinusoidal
contributions to the density on the right-hand side of equation (26).

The second term on the right-hand side of equation (65) is mildly singular as $E_R \to 0$.
Therefore, the requirement that the total density of
stars in the six-dimensional phase space must be non-negative implies that
$$
\lambda(\bmath{m},\bmath{G}_\perp,G_0) \exp({\rm i} \theta_0) \le 0 , \eqno\stepeq
$$
where it will be remembered
that $\lambda(\bmath{m},\bmath{G}_\perp,G_0) \exp({\rm i} \theta_0)$
is a real quantity (see eq. [62]).

The singularity of $f_R (H_R,\bmath{G}_\perp)$ at $E_R = 0$ implies that the resonant stars tend to
concentrate near the separatrix defined by equation (44) ({\rm cf.} eq.[60]).

If $\lambda(\bmath{m},\bmath{G}_\perp,G_0) = 0$, then the oscillation described here is a
slightly nonlinear counterpart of a mode of the Vlasov type.  According to equation (58),
resonant stars are not required to support the oscillation of the system.  Moreover, the mode
satisfies the characteristic equation ${\rm det} |M (\sigma,{\bmath{\alpha}},{\bmath{\beta}})| = 0$
as described in Section 2.3.2.  According to equation (65), the density of stars in the resonance
region of the phase space is uniform in that case.

\subsection{A constraint on modes of the van Kampen type}

Inequality (66) constrains the possible frequencies of a stationary oscillation described by
equation (58).  In order to derive the constraint, we multiply each term in equation (58) by
$\mu^* (\bmath{\alpha},\sigma)$, sum over the values of $\bmath{\alpha}$, and reduce the resulting
equation with the aid of equations (16), (22), and ((41).  Finally, we make use of the condition
$\cos (\theta_0 + \phi) = - \eta$ governing the value of $\theta_R = \theta_0$ at a stable
fixed point of the Hamiltonian represented by equation (42) (see also equations [40]).
We obtain
$$
{1 \over 2} (2\upi)^3 \eta \epsilon m_*
\int {\rm d}G_2 {\rm d}G_3 |V_1 (\bmath{m},\bmath{m}G_0 + \bmath{G}_\perp)|
 \lambda(\bmath{m},\bmath{G}_\perp,G_0) \exp({\rm i} \theta_o)
= {1 \over 2} \epsilon
\Big[ 1 - \sum_\bmath{n}
(2 \upi)^3 m_* P \int {\rm d} \bmath{J}
{ { | V_1 (\bmath{n},\bmath{J}) |^2 }
\over { \sigma - \bmath{n} \cdot \bmath{\omega} (\bmath{J}) } }
 \bmath{n} \cdot { { \upartial f_0 } \over { \upartial \bmath{J} } } \Big] .
 \eqno\stepeq
$$
It follows from inequality (66) that the product of $\eta$ and the left hand side of
equation (67) must be negative or zero.  Therefore,
$$
\eta \Big[ 1 - \sum_\bmath{n}
(2 \upi)^3 m_* P \int {\rm d} \bmath{J}
{ { | V_1 (\bmath{n},\bmath{J}) |^2 }
\over { \sigma - \bmath{n} \cdot \bmath{\omega} (\bmath{J}) } }
 \bmath{n} \cdot { { \upartial f_0 } \over { \upartial \bmath{J} } } \Big] \le 0 .
 \eqno\stepeq
$$

The equality holds in condition (68) in the case that
$\lambda(\bmath{m},\bmath{G}_\perp,G_0) = 0$.  In other words, condition (68) implies
that the continuous spectrum of frequencies of the modes of the van Kampen type is bounded by
the frequency of a mode of the Vlasov type.  The continuous spectrum
might be bounded either from above or from below, depending on the sign of $\eta$
(see equations [40]) and on the structure of the integrand in condition (68).

The constraint imposed by condition (68) on the frequencies of plane waves in an infinite
homogeneous stellar system is described in Appendices B and C.

\section{Nonlinear oscillations of a galaxy}

In principle, the series in powers of $\epsilon$ constructed in Section 3 for the representation
of slightly nonlinear oscillations could be extended to include terms of higher orders in
$\epsilon$.  However, there does not appear to be any outstanding issue of principle that
would require the investigation of terms of higher orders.  Moreover, the construction of
terms of higher orders in the series representation of slightly nonlinear oscillations of a
galaxy would rapidly become complicated.  Therefore, it seems preferable to find alternative
ways to satisfy equations (26), (27), and (46) for fully nonlinear oscillations without
resorting to the series solutions started in Section 3.

For a particular model of a radial oscillation of a spherically symmetric galaxy, Louis
and Gerhard (1988) have solved equations (26), (27), and (46) with the aid of a generalized
version of Schwarzschild{\rq}s (1979) numerical method for the construction of models of
galaxies.  Specifically, they have made use of Lucy{\rq}s (1974) method in order to solve
the discrete set of equations with which Schwarzschild replaces the integral equation (46).

In their discussion and interpretation of their results, Louis and Gerhard touch upon certain
features of their model which are reproduced more generally in the present work.  By
investigating the arrangement of resonant orbits in the phase space and exhibiting the
solution of Schwarzschild{\rq}s equations for the distribution of stars on the resonant
orbits, they demonstrate that the self-consistency of the oscillation is sustained by an
appropriate population of resonant orbits with stars.  In that connection, they present
and analyse the pendulum model described in Section 3.3 above in the form appropriate to
conditions of spherical symmetry.  Their surfaces of section exhibit the arrangement of
the resonance regions of the phase space as closed chains of islands as described at the
end of Section 3.3 above.  Finally, Louis and Gerhard remark that resonant {\lq}orbits
near the unstable closed resonant orbits support the perturbation.{\rq}\/
This accords with the more general result found at the end of Section 3. 5 above that
the resonant stars tend to concentrate near the separatrix defined by equation (44).

\section{Discussion and concluding remarks}

The work described in this paper generalizes the theory of van Kampen modes in plasmas to
the gravitational case of modes of oscillation in inhomogeneous stellar systems,
and it does so with the aid of a
modified formulation of the matrix method of Kalnajs.  The present construction of normal
modes, particularly modes with continuous spectra of real frequencies, complements
more conventional, initial-value treatments of small perturbations in galaxies.

We have found that the connection between van Kampen modes and nonlinear plasma waves
carries over to a connection in stellar dynamics between linear modes of stationary
oscillation and slightly nonlinear, stationary oscillations.  That connection also
bridges the gap between the linear theory of small perturbations
in stellar systems and nonlinear models of oscillating systems of the kind considered by
Louis and Gerhard (1988).  The slightly nonlinear theory of stationary oscillations
described in Section 3 should provide an analytical foundation for more general and more
extensive numerical investigations along the lines described by Louis and Gerhard.

The treatment of stationary oscillations described in this paper is quite general,
and it has wide applications.  The principal restriction underlying the construction of
linear oscillations in Section 2 and slightly nonlinear oscillations in Section 3 is
that the stellar motions in the unperturbed gravitational field of the system must be
separable.  The KAM theorems imply that the series in powers of the small parameter
$\epsilon$ developed in Section 3 are a valid approximation, for sufficiently small
values of $\epsilon$, in the sense that the perturbed motions of stars remain
separable for most initial conditions.  Thus, the theories of stationary oscillations of
stellar systems described in this paper apply to spherically symmetric systems and to
axisymmetric and triaxial St{\"a}ckel systems.  The KAM theorems also imply that the present
representation of stationary oscillations in galaxies would apply, at least approximately,
to systems in which the unperturbed stellar motions are only approximately separable.
This would include applications to systems such as slowly-rotating, triaxial systems.

On the other hand, for sufficiently large values of $\epsilon$, the overlap of
resonances will lead to chaotic behavior of the relevant stellar orbits (Chirikov
1979).  Such chaotic behavior would be expected to inhibit stationary oscillations at
sufficiently large amplitudes.  Inasmuch as orbits near the separatrices that form the
boundaries of resonance regions would be the first to become chaotic, the onset of
chaos in the oscillating systems considered here would probably be enhanced
by the tendency of resonant stars to concentrate near the separatrices.

Although the oscillations described in this paper are sinusoidal, it appears that
more general periodic oscillations would arise in a more fully nonlinear
treatment.  Louis and Gerhard (1988) showed this to be the case for their model, but
they argued that the relevant nonlinearities were small enough that they could
approximate the oscillation of the model as sinusoidal.

It is an important result of the present investigation that nonlinear considerations
impose the constraint represented by inequality (68) on the continuous spectrum of modes
of the van Kampen type and that a mode of the Vlasov type plays a special role
in this connection.  The mode of the Vlasov type is the limiting case
of modes of the van Kampen type in which resonant stars do not contribute to the
self-consistency of the oscillation.  Moreover, according to inequality (68),
the continuous spectrum of frequencies of modes of the van Kampen type is bounded
by the frequency of the mode of the Vlasov type (see particularly the example described
in the appendices of this paper).  These results, which seem not to be
recognized in the literature of plasma physics, raise an important issue regarding
the connection between Landau{\rq}s and van Kampen{\rq}s treatments of plasma
oscillations.  The point of inequality (68) is that the excluded modes of the van
Kampen type are unphysical in the sense that involve negative densities of resonant
stars in the phase space.  One might ask if unphysical van Kampen modes can be
included in superpositions of modes that would represent solutions of the initial
value problem of the Landau type.  There is certainly no mathematical objection to
the use of unphysical modes in a superpositon, and there can be no physical objection
if the superposition preserves positive densities of stars everywhere in the phase
space.  Thus, inequality (68) serves mainly to identify those modes of the van
Kampen type which, in isolation, may be regarded as physical perturbations.

Stix (1992, see particularly Chapters 7 and 8) has emphasized that the connection
between van Kampen modes and BGK waves validates the consideration of the van Kampen
modes individually as distinct modes of oscillation in a plasma.  The results
described in this paper serve likewise to validate the consideration of stationary
oscillations of galaxies.  It remains, however, to identify mechanisms that could
excite stationary oscillations of galaxies in nature.  That is a question
for future research.  Nevertheless, one might speculate that, in the formation or growth
of a galaxy through mergers, a stationary oscillation could be excited as an accreted
fragment becomes trapped in a resonance.  And one might ask if dynamical friction acting
on an accreted fragment could contribute to such trapping.

\appendix

\section{Van Kampen modes in an infinite, homogeneous system}

Perturbations of a system in which the density is uniform and
the distribution function $f_0 ({\bmath{v}})$ depends only on the stellar velocities
provide an illustration of the matrix method described in the body of this paper.
The model is the gravitational counterpart of the theory of plane waves in a homogeneous
plasma (van Kampen 1955; Stix 1992).  The evolution of small perturbations in an
infinite, homogeneous stellar system has been
treated as an initial-value problem by Lynden-Bell (1962) and Sweet (1963), and the
complementary normal-mode analysis is discussed briefly by Binney and Tremaine (1987).

In order to construct a suitable set of action-angle variables and a suitable biorthonormal
set of densities and potentials, we employ a \lq{box normalization}\rq\/
and concentrate on the unit cell $0 \le x_i \le L$ $(i = 1,2,3)$.
The action-angle variables of the unperturbed motion of a star are
$$
{\bmath{J}} = {{\bmath{v}} L \over {2 \upi}} \quad {\rm and} \quad
{\bmath{w}} = {{2 \upi {\bmath{x}}} \over L } . \eqno\stepeq
$$
The Hamiltonian and the frequencies of the unperturbed motion are
$$
H_0({\bmath{v}}) = {1 \over 2} |{\bmath{v}}|^2 = {1 \over 2} \Bigl( {{2 \upi} \over L}\Bigr)^2 |{\bmath{J}}|^2
= H_0 ({\bmath{J}}) 
\quad \hbox{ and } \quad
{\bmath{\omega}} ({\bmath{J}}) = {{\upartial H_0} \over {\upartial {\bmath{J}}}}
= \Bigl( {{2 \upi} \over L}\Bigr)^2 {\bmath{J}} = \Bigl( {{2 \upi} \over L}\Bigr) {\bmath{v}} ,
\eqno\stepeq
$$
respectively.  The members of the basis set are of the form
$$
V_1 ({\bmath{x}},{\bmath{\kappa}}) = V_1 ({\bmath{\kappa}}) \exp ({\rm i} {\bf k} \cdot {\bmath{x}})
\quad \hbox{ and } \quad
\rho_1 ({\bmath{x}},{\bmath{\kappa}}) = \rho_1 ({\bmath{\kappa}}) \exp ({\rm i} {\bf k} \cdot {\bmath{x}})
= - {{|{\bf k}|^2} \over {4 \upi G}}V_1 ({\bmath{\kappa}}) \exp ({\rm i} {\bf k} \cdot {\bmath{x}}),
\eqno\stepeq
$$
where ${\bmath{\kappa}} = (\kappa_1,\kappa_2,\kappa_3)$ is a set of three integers and
${\bf k} = 2 \upi {\bmath{\kappa}} / L$.  Thus the basis functions satisfy periodic boundary
conditions on the unit cell.  In the orthogonality relations, equations (13), the unit cell
is the volume of integration.  In particular, the normalization of these bases gives
$$
{{|{\bf k}|^2}\over {4 \upi G}} |V_1 ({\bmath{\kappa}})|^2 L^3 = 1 . \eqno\stepeq
$$

We must reduce equations (11) and (17) with the aid of equations (A1)-(A4).
Upon writing equation (15) explicitly for this purpose, we find that
$$
 V_1 ({\bmath{n}},{\bmath{J}},{\bmath{\beta}}) = \delta ({\bmath{n}},{\bmath{\beta}}) V_1 ({\bmath{\beta}}) , \eqno\stepeq
$$
where $\delta ({\bmath{n}},{\bmath{\beta}})$ is a Kronecker delta.
When we simplify equation (17) with the aid of equation (A5)), we find
that the modes must be plane waves
$\exp ({\rm i} {\bf k} \cdot {\bmath{x}} - {\rm i} \sigma t) =
\exp ({\rm i} {\bmath{\kappa}} \cdot {\bmath{w}} - {\rm i} \sigma t)$.  Thus, in equations (11)
and (17), we have $ \mu({\bmath{\alpha}},\sigma) = 0$ and
$ \lambda ({\bmath{\alpha}}, {\bmath{\varpi}}, \sigma) = 0$ $({\bmath{\alpha} } \ne {\bmath{\kappa}})$
and, in virtue of equation (22), $ \mu({\bmath{\kappa}},\sigma) = 1$.  This reduces each
of the trigonometric series in equation (11) to a single term in
$\exp ({\rm i} {\bmath{\kappa}} \cdot {\bmath{w}})$.

The condition ${\sigma - {\bmath{n}} \cdot {\bmath{\omega}} ({\bmath{J}})} = 0$ in the case that
${\bmath{n}} = {\bmath{\kappa}}$ reduces to  ${\sigma - {\bf k} \cdot {\bmath{v}}} = 0$.  By resolving
${\bmath{v}}$ into a component $ v_k $ parallel to ${\bf k}$ and a component ${\bmath{v}}_\perp$
perpendicular to ${\bf k}$, we can reduce the conditions
${\bmath{J}} = {\bmath{J}}_R ({\bmath{n}},{\bmath{\varpi}},\sigma)$ to $ v_k = \sigma / |{\bf k}|$ and
${\bmath{v}}_\perp = {\bmath{\varpi}}$.

Equation (11) reduces to 
$$
f_1 ({\bmath{x}},{\bmath{v}},t )
= - P { {V_1 ({\bmath{\kappa}})} \over 
{\sigma - {\bf k} \cdot {\bmath{v}} }}
{\bf k} \cdot {{\upartial f_0} \over {\upartial {\bmath{v}}}}
 \exp ({\rm i} {\bf k} \cdot {\bmath{x}} - {\rm i} \sigma t)
+ \Bigr({{2 \upi} \over L}\Bigl)^3
\lambda  ({\bmath{\kappa}}, {\bmath{v}}_\perp , \sigma)
\delta (v_k - \sigma / |{\bf k}|)
\exp ({\rm i} {\bf k} \cdot {\bmath{x}} - {\rm i} \sigma t) .  \eqno\stepeq
$$
Similarly, equation (17) reduces to 
$$
1 - {{4 \upi G m_*} \over {|{\bf k}|^2}} P { \int {\rm d} {\bmath{v}}
{1 \over {\sigma - {\bf k} \cdot {\bmath{v}} }}
{\bf k} \cdot {{\upartial f_0} \over {\upartial {\bmath{v}}}}}
= - (2\upi)^3 m_* V_1^* ({\bmath{\kappa}}) \int {\rm d} {\bmath{v}_\perp}
\lambda  ({\bmath{\kappa}}, {\bmath{v}}_\perp , \sigma) . \eqno\stepeq
$$
Equations (A6) and (A7) describe the gravitational analogue of van Kampen\rq{s}
solution for the electrostatic modes of oscillation of a homogeneous plasma
(van Kampen 1955).

For a mode in which the frequency $\sigma$ is real, equation (A7) fixes the value of
the integral of $\lambda  ({\bmath{\kappa}}, {\bmath{v}}_\perp , \sigma)$ over
$\bmath{v}_\perp$ on the right-hand side, and there is no dispersion relation.
Of course, the exceptional case is one in which $\lambda  ({\bmath{\kappa}}, {\bmath{v}}_\perp , \sigma)$
vanishes, and equation (A7) reduces to the gravitational counterpart of Vlasov{\rq}s (1945) dispersion
relation.

When $|\bmath{k}| \le k_J$, where $k_J$ is the Jeans wave number, and
$\sigma - {\bf k} \cdot {\bmath{v}} \ne 0$, equation (A7) reduces to
$$
1 - {{4 \upi G m_*} \over {|{\bf k}|^2}} { \int {\rm d} {\bmath{v}}
{1 \over {\sigma - {\bf k} \cdot {\bmath{v}} }}
{\bf k} \cdot {{\upartial f_0} \over {\upartial {\bmath{v}}}}} = 0 , \eqno\stepeq
$$
a dispersion relation which admits unstable Jeans modes (Binney and Tremaine 1987).
If the velocity distribution $f_0 (v)$ is Maxwellian, then 
${k_J}^2 = 12\upi G \rho/\langle v^2 \rangle$, where $\langle v^2 \rangle$ is the
mean-square velocity of a star in three dimensions.

\section{BGK waves in an infinite, homogeneous system}

In this appendix, we construct a gravitational counterpart of a BGK wave (Bernstein, et al. 1957)
in an infinite, homogeneous stellar system as an example of a slightly nonlinear oscillation
of the kind described in Section 3.

As in the case of a van Kampen mode of the kind described in Appendix A, the perturbation
of the system consists of a single plane wave.  Symbols not defined in what follows have
the meanings assigned in the body of the paper or in Appendix A.  Thus the density of the
system is of the form
$$
\rho ({\bmath{x}},t) = \rho_0 
+ {1 \over 2} \epsilon \rho_1 ({\bmath{\kappa}})
\exp ({\rm i} {\bf k} \cdot {\bmath{x}} - {\rm i} \sigma t)
+ {1 \over 2} \epsilon \rho_1^* ({\bmath{\kappa}})
\exp (-{\rm i} {\bf k} \cdot {\bmath{x}} + {\rm i} \sigma t) , \eqno\stepeq
$$
where $\rho_0$ is the constant density of the system in the absence of the wave,
and the gravitational force per unit mass acting on a body is derived from the
potential
$$
V ({\bmath{x}},t) =
{1 \over 2} \epsilon V_1 ({\bmath{\kappa}})
\exp ({\rm i} {\bf k} \cdot {\bmath{x}} - {\rm i} \sigma t)
+ {1 \over 2} \epsilon V^* ({\bmath{\kappa}})
\exp (-{\rm i} {\bf k} \cdot {\bmath{x}} + {\rm i} \sigma t) . \eqno\stepeq
$$
Here, the wave is described in terms of a single member of the biorthonormal basis
set of potentials and densities represented in equations (A3).  The Hamiltonian
is now of the form
$$
H = {1 \over 2} | \bmath{v} |^2
+ {1 \over 2} \epsilon V_1 ({\bmath{\kappa}})
\exp ({\rm i} {\bf k} \cdot {\bmath{x}} - {\rm i} \sigma t)
+ {1 \over 2} \epsilon V^* ({\bmath{\kappa}})
\exp (-{\rm i} {\bf k} \cdot {\bmath{x}} + {\rm i} \sigma t) . \eqno\stepeq
$$

In virtue of equations (A1) and (A2) the resonant condition reduces to
$\sigma - \bmath{k} \cdot \bmath{v} = 0$ in this case.  The Hamiltonian contains no
non-resonant terms.  Therefore, the canonical perturbation
theory described in Section 3.2 simplifies so that the transformation equations 
(A1), (31), (32), (35), and (36) reduce to
$$
\eqalign{
{{\bmath{v} L} \over {2 \upi}} & = \bmath{J} = \bmath{I}
= \bmath{\kappa} G_R + \bmath{G}_\perp , \cr
\theta_R &= \bmath{\kappa} \cdot \bmath{\psi} - \sigma t = \bmath{\kappa} \cdot \bmath{w} - \sigma t
= \bmath{k} \cdot \bmath{x} - \sigma t , \quad
\theta_2 = \psi_2 = w_2 = {{2 \upi x_2} \over L} ,  \quad {\rm and} \quad
\theta_3 = \psi_3 = w_3 = {{2 \upi x_3} \over L} , \cr } \eqno\stepeq
$$
and we are left with the Hamiltonian
$$
H_R (\theta_R,G_R ,\bmath{G}_\perp)
 = {1 \over 2} \Big({{2 \upi} \over L}\Big)^2
|\bmath{\kappa} G_R + \bmath{G}_\perp|^2  - \sigma G_R
+ {1 \over 2} \epsilon V_1 (\bmath{\kappa}) \exp ( {\rm i} \theta_R)
+ {1 \over 2} \epsilon V_1^* (\bmath{\kappa}) \exp ( - {\rm i} \theta_R)
  \eqno\stepeq
$$
(see eqs. [37] and [A5] and note that we are now replacing $\bmath{m}$ with
$\bmath{\kappa}$).

In the reduction of the right-hand side of equation (B5) to the Hamiltonian of a pendulum
model, we find that equations (40) reduce to $\eta = +1$ and $\mu = |\bmath{k}|^2 $ in virtue of
the second of equations (A2) and the first of equations (B4).  As in Appendix A, we introduce
components $v_k$ and $\bmath{v}_\perp$ of $\bmath{v}$ parallel and perpendicular, respectively,
to the wave vector $\bmath{k}$.  With the aid of the first of equations (B4), we find that
$$
v_k = kG_R + {{2 \upi} \over {k L}} \bmath{k} \cdot \bmath{G}_\perp  \quad {\rm and} \quad
\bmath{v}_\perp = {{2 \upi} \over L} \Big[ \bmath{G}_\perp
- { {\bmath{k} (\bmath{k} \cdot \bmath{G}_\perp ) } \over k^2} \Big] .  \eqno\stepeq
$$
Thus for the Hamiltonian of the pendulum model in the case at hand, we replace equation (43) with
$$
H_R (\theta_R,G_R ,\bmath{G}_\perp) = {1 \over 2} v_0^2 + {1 \over 2} |\bmath{v}_\perp |^2
- \sigma {v_0 \over k} + \sigma { {2 \upi} \over L} {{\bmath{k} \cdot \bmath{G}_\perp }
\over k^2}
+ {1 \over 2} (v_k - v_0)^2 - \epsilon |V_1 (\kappa)| \cos [k(x_R - x_0)] , \eqno\stepeq
$$
where $v_0 = \sigma / k$ is the phase velocity of the wave. Here, we have also defined $x_R$
and $x_0$ by writing $\theta_R = k x_R - \sigma t$ and $\theta_0 = k x_0 - \sigma t$, respectively,
where it will be remembered that $\theta_0$ is the value of the angle variable $\theta_R$ at the
stable fixed point of the Hamiltonian.

The distribution function for the stars trapped in the wave, { \rm i. e. }, for the resonant stars, is now
$$
f_R (H_R,\bmath{G}_\perp)
= f_0 (\bmath{v}_0)
- \Big({1 \over 2}\Big)^{1/2}
{ { \lambda(\bmath{\kappa}) \exp({\rm i} \theta_0) } \over
{ | \kappa_1 | \upi |V_1 (\bmath{\kappa})|}}
[ \epsilon |V_1 (\bmath{\kappa})| (-E_R)^{-1/2}
-2(-E_R)^{1/2}]  ,  \eqno\stepeq
$$
where $\bmath{v}_0 = (\bmath{k} v_0/k) + \bmath{v}_\perp$ and
$$
E_R = {1 \over 2} (v_k - v_0)^2 - \epsilon |V_1 (\kappa)| \{ 1 + \cos [k(x_R - x_0)] \} 
\eqno\stepeq
$$
(see eqs.[60] and [65]).

The distribution function for the untrapped stars, { \rm i. e. }, for the nonresonant stars, is
$$
f_{NR} (\bmath{x},\bmath{v},t)
= f_0 (\bmath{v})
- {1 \over 2} \epsilon \Big[ { { V_1 ({\bmath{\kappa}})
\exp ({\rm i} {\bmath{k}} \cdot {\bmath{x}} - {\rm i} \sigma t) } \over
{ \sigma - \bmath{k} \cdot \bmath{x} } }
+ { V_1^* ({\bmath{\kappa}})
\exp ( - {\rm i} {\bmath{k}} \cdot {\bmath{x}} + {\rm i} \sigma t) \over
{ \sigma - \bmath{k} \cdot \bmath{x} } } \Big]
\bmath{k} \cdot {{\upartial f_0} \over {\upartial \bmath{v}}}
 \eqno\stepeq
$$
(see eq.[48]).

Finally, inasmuch as the perturbation of the potential consists of a single basis function
$V_1 ({\bmath{\kappa}}) \exp ({\rm i} {\bf k} \cdot {\bmath{x}} - {\rm i} \sigma t)$,
equation (58) expressing the self-consistency of the wave reduces to
$$
{1 \over 2} (2\upi)^3 \epsilon m_*
\int {\rm d}G_2 {\rm d}G_3 V_1^* (\bmath{\kappa}) \lambda(\bmath{\kappa},\bmath{G}_\perp,G_0)
= - {1 \over 2} \epsilon
\Big[ 1 -  { {4 \upi G m_* } \over k^2} P \int {\rm d} \bmath{v}
{ 1 \over { \sigma - \bmath{k} \cdot \bmath{v} } }
 \bmath{k} \cdot { { \upartial f_0 } \over { \upartial \bmath{v} } } \Big] \eqno\stepeq
$$
in virtue of equations (A1) and (A4).  Equation (B11) is to be compared with equation (A8).

For a plane wave described by equation (B11), inequality (68) reads.
$$
1 -  { {4 \upi G m_* } \over k^2} P \int {\rm d} \bmath{v}
{ 1 \over { \sigma - \bmath{k} \cdot \bmath{v} } }
 \bmath{k} \cdot { { \upartial f_0 } \over { \upartial \bmath{v} } } \le 0 .
 \eqno\stepeq
$$
Where the equality holds, this condition gives the gravitational counterpart of
Vlasov{\rq}s dispersion relation for a plasma wave.  The manner in which the mode
of the Vlasov type bounds the allowed, continuous spectrum of modes of the van Kampen
type is illustrated in the appendix that follows.

\section{An example of an allowed spectrum of van Kampen modes}

We define a one-dimensional velocity distribution
$$
F_0 (v_k) = \int f_0 (\bmath{v}_0 ) {\rm d} \bmath{v}_\perp \eqno\stepeq
$$
for the unperturbed system, and we let
$$
F_0 (v_k) = { {3 j \rho_0} \over {4 m_*}} (1 - j^2 v_k^2) \qquad (v_k^2 \le j^{-2} ) 
\quad \hbox{ and } \quad F_0 (v_k) = 0 \qquad (v_k^2 \ge j^{-2} ) ,  \eqno\stepeq
$$
where $j$ is a constant.

For this model, the integral in inequality (B12) is readily evaluated and the inequality
brought to the form
$$
{\sigma^2 \over {6 \upi G \rho_0}} \le 
{1 \over y^3} \Big[ 2y - \ln \Big( {{y+1} \over {y-1}} \Big) \Big] , \eqno\stepeq
$$
where $y = k/(\sigma j)$ and, without loss of generality, we let $ \sigma \ge 0$.

For an assigned value of $y$, the right-hand side of inequality (C3) sets an upper bound
$\sigma_{\rm max}$ on the magnitude of $\sigma$.  Thus, $\sigma = \sigma_{\rm max}$ at a wave
number $k = \sigma j y$.
We have $\sigma = \sigma_{\rm max}$ when the equality holds in conditions (B12) and (C3).
In other words, $\sigma_{\rm max}$ satisfies the gravitational counterpart of the
Vlasov{\rq}s (1945) dispersion relation.

Figure A1 is a plot of $\sigma_{\rm max}$ against $k$ for a plane wave in the model that we
are considering.  Here, we  measure frequencies in the unit $(6 \upi G \rho_0)^{1/2}$ and
wave numbers in the unit $(6 \upi G \rho_0 j^2 )^{1/2}$.  The frequencies in the continuous
spectrum of the van Kampen modes which are allowed by inequality (C3) fill the
region under the curve in Figure A1.  Note that $\sigma_{\rm max} \to 0$ as
$k \to k_J = (12 \upi G \rho_0 j^2 )^{1/2}$, where $k_J$ ($=2^{1/2}$ in the adopted system
of units) is the Jeans wave number for the model.  Evidently, the allowed modes of the
van Kampen type occur in this model only at wave numbers such that $ k \le k_J$ and the
system is gravitationally unstable.

\beginfigure*{A1}
\vskip 67mm
\caption{{\bf Figure A1.} The allowed real frequencies in the continuous spectrum of
modes of the van Kampen type in an infinite, homogeneous stellar system.  For a plane
wave, the curve represents the solution of a dispersion relation of the Vlasov (1945)
type for the dependence of the frequency $\sigma$, measured in the unit
$(6 \upi G \rho_0)^{1/2}$, on the wave number $k$, measured in the unit
$(6 \upi G \rho_0 j^2 )^{1/2}$.  The allowed frequencies of the van Kampen modes
fill the region under the curve representing the Vlasov mode.}
\endfigure

\bye